\documentclass[aps,prc,floatfix,groupedaddress,showpacs,amsfonts]{revtex4}
\usepackage{bm}
\usepackage{epsfig}
\usepackage{amsmath}

\newcommand*{\Rcu}{{\mathcal R}}

\begin{document}

\title{Coulomb renormalization and ratio of proton and neutron asymptotic normalization coefficients for mirror nuclei}

\author{A.\,M.~Mukhamedzhanov}
\affiliation{Cyclotron Institute, Texas A\&M University,
College Station, TX 77843}

\begin{abstract}
Asymptotic normalization coefficients (ANCs) are fundamental nuclear constants playing important role in nuclear reactions, nuclear structure and nuclear astrophysics. In this paper the physical reasons of the Coulomb renormalization of the ANC are addressed. Using Pinkston-Satchler equation the ratio for the proton and neutron ANCs of mirror nuclei is obtained in terms of the Wronskians from the radial overlap functions and regular solutions of the two-body Schr\"odinger equation with the short-range interaction excluded. This ratio allows one to use microscopic overlap functions for mirror nuclei in the internal region, where they are the most accurate, to correctly predict the ratio of the ANCs for mirror nuclei, which determine the amplitudes of the tails of the overlap functions. Calculations presented for different nuclei demonstrate the Coulomb renormalization effects and independence of the ratio of the nucleon ANCs for mirror nuclei on the channel radius. This ratio is valid both for bound states and resonances. One of the goals of this paper is to draw attention on the possibility to use the Coulomb renormalized ANCs rather than the standard ones especially when the standard ANCs are too large.

\end{abstract}


\maketitle

\section{Introduction}
The asymptotic normalization coefficient (ANC) is a fundamental nuclear characteristics of bound states \cite{blokh77,blokhintsev84} and resonances (the partial resonance width is also expressed in terms of the ANC) \cite{muk77,muktr99}  playing an important role in nuclear reaction and structure physics. The ANSs determine the normalization of the peripheral part of transfer reaction amplitudes \cite{blokh77,blokhintsev84} and overall normalization of the peripheral radiative capture processes \cite{muktim90,muk90,muk2001,muk2011}.  
In the R-matrix approach the ANC determines the normalization of the external nonresonant radiative capture amplitude and the channel radiative reduced width amplitude \cite{muktr99,tang,kroha2011}.

The ANC enters the theory in two ways \cite{blokh77}. In the scattering theory the residue at the poles of the elastic scattering $S$ matrix corresponding to bound states \cite{kramers,perelomov} or resonances \cite{muk77} can be expressed in terms of the ANC:
\begin{align}
S_{l_{B}\,j_{B};l_{B}\,j_{B}}^{J_{B}} \xrightarrow{k \to k_{aA}^{p}}\,\frac{A_{l_{B}\,j_{B}}}{k - k_{aA}^{p}}    
\label{elscatSm1}
\end{align}
with the residue
\begin{align}
A_{l_{B}\,j_{B}}^{J_{B}}= -i^{2\,l_{B} + 1}\,e^{i\,\pi\,\eta_{aA}^{p}}\,\big(C^{B}_{aA\,l_{B}\,j_{B}\,J_{B}}\big)^{2},
\label{residue1}
\end{align}
where $\,C^{B}_{aA\,l_{B}\,j_{B}\,J_{B}}$ is the ANC for the virtual or real decay $B \to a + A$ in the channel with the relative orbital angular momentum $\,\,l_{B}$ of $a$ and $A$, the total angular momentum $\,\,j_{B}$ of $a$ and total angular momentum $\,\,J_{B}$ of the system $a + A$. If $B=(a\,A)$ is a bound state, 
\begin{align}
\eta_{aA}^{p} \equiv \eta_{aA}^{bs}=\frac{Z_{a}\,Z_{A}\,e^{2}\,\mu_{aA}}{\kappa_{aA}^{B}}
\label{eta01}
\end{align}
is the Coulomb parameter for the bound state $B=(a\,A)$, $k_{aA}^{p}= i\,\kappa_{aA}^{B}$, $\,\,\kappa_{aA}^{B}= \sqrt{2\,\mu_{aA}\,\varepsilon_{aA}^{B}}$ is the bound-state wave number, $\,\varepsilon_{aA}^{B}= m_{a} + m_{A} - m_{B}$ is the binding energy for the virtual decay $B \to a + A$, $Z_{i}\,e$ and $m_{i}$ is the charge and mass of particle $i$, and $\mu_{aA}$ is the reduced mass of $a$ and $A$. Throughout the paper I use the system of units in which $\hbar=c=1$.  Note that the singling out the factor
$e^{i\,\pi\,\eta_{aA}^{p}}$ in the residue makes the ANC for bound states real (see Sect. \ref{wronskian}).

If $B$ is a resonance, that is the decay $B \to a + A$ is real, then 
\begin{align} 
&\eta_{aA}^{p} =i\,\eta_{aA}^{R}, \qquad   \eta_{aA}^{R}=\,\frac{Z_{a}\,Z_{A}\,e^{2}\,\mu_{aA}}{k_{aA\,(R)}}
\label{etaR1}
\end{align}
is the Coulomb parameter for the resonance state with complex relative momentum $k_{aA}^{p}=k_{aA\,(R)}$, $\,\,k_{aA\,(R)}=\sqrt{2\,\mu_{aA}\,E_{aA\,(R)}}$, $\,\,E_{aA\,(R)}$ is the resonance energy in the system $a + A$ of the resonance state $B=(aA)$. Eqs (\ref{elscatSm1}) and (\ref{residue1}) being universal valid for both bound state poles and resonances \cite{muk77} provide the most general and model-independent definition of the ANC. 

In the case of the Breit-Wigner resonance  (${\rm Im}k_{aA\,(R)} << {\rm Re}k_{aA\,(R)}= k_{aA\,(R)}^{0}$) the residue of the elastic scattering $S$-matrix element in terms of the resonance width is  
\begin{align}
A_{l_{B}j_{B}}^{J_{B}}= -\,ie^{2\,i\,\delta_{l_{B}\,j_{B}\,J_{B}}^{p}(k_{aA}^{0})}\,\frac{\mu_{aA}}{k_{aA\,(R)}^{0}}\,\Gamma_{aA\,l_{B}j_{B}J_{B}},
\label{residreswidth1}
\end{align}
where $\delta_{l_{B}\,j_{B}\,J_{B}}^{p}(k_{aA}^{0})$ is the potential (non-resonance) scattering phase shift at the real resonance relative momentum $k_{aA\,(R)}^{0}$.
Meantime Eq.(\ref{residue1}) in the Breit-Wigner resonance case takes the form 
\begin{align}
A_{l_{B}\,j_{B}}^{J_{B}}= -i^{2\,l_{B} + 1}\,e^{-\,\pi\,\eta_{aA}^{(0)}}\,\big(C^{B}_{aA\,l_{B}\,j_{B}\,J_{B}}\big)^{2},
\label{residueres1}
\end{align}
where $\eta_{aA}^{(0)}= Z_{a}\,Z_{A}\,e^{2}\,\mu_{aA}/k_{aA\,(R)}^{0}$.
Then the ANC and the partial width of the resonance are related by
\begin{align}
\big(C^{B}_{aA\,l_{B}\,j_{B}\,J_{B}}\big)^{2}= (-1)^{l_{B}}\,e^{\pi\,\eta_{aA}^{(0)}}\,e^{i\,2\,\delta_{l_{B}\,j_{B}\,J_{B}}^{p}(k_{aA(R)}^{0})}\,\frac{\mu_{aA}\,\Gamma_{aA\,l_{B}\,j_{B}\,J_{B}}}{k_{aA(R)}^{0}}.
\label{ANCGamma1}
\end{align}

In the case of the subthreshold resonance (the resonance which is a bound state in the entry channel and a resonance in the exit channel) the connection between the ANC and the partial width, calculated at the relative momentum of particles $a$ and $A$ $\,\,k_{aA}$, is given by   \cite{muktr99}
\begin{align}
\Gamma_{aA\,l_{B}\,j_{B}\,J_{B}}(k_{aA})= P_{l_{B}}(k_{aA}\,R_{aA})\,\frac{W_{-\eta_{aA}^{bs},\,l_{B} + 1/2}^{2}(2\,\kappa_{aA}^{B}\,R_{aA})}{\mu_{aA}\,R_{aA}}\,\big(C^{B}_{aA\,l_{B}\,j_{B}\,J_{B}}\big)^{2},
\label{gammasubthrshresANC1}
\end{align}
where $P_{l_{B}}(k_{aA}\,R_{aA})$ is the barrier penetrability, $W_{-\eta_{aA}^{bs},\,l_{B} +1/2}(2\,\kappa_{aA}^{B}\,R_{aA})$ is the Whittaker function for the bound state $B=(aA)$ and $R_{aA}$ is the channel radius. 
Correspondingly, the reduced width is determined by
\begin{align}
\gamma _{aA\,{l_B}\,{j_B}\,J_{B}}^2 = \frac{{W_{ - {\eta _{aA}^{bs}},\,{l_B}\, + \,1/2}^2(2{\kappa _{aA}^{B}}{R_{aA}})\,}}{{2{\mu _{aA}}{R_{aA}}}}\,\big(C^{B}_{aA\,{l_B}\,{j_B}\,J_{B}}\big)^2.
\label{redwidth1}
\end{align}
Note that in the $R$-matrix method the peripheral part of the radiative capture amplitude is expressed in terms 
of the reduced width rather than the ANC. However, the reduced width is model dependent, because it depends on the channel radius $R_{aA}$, while the ANC is not. 

From other side, in the Schr\"odinger formalism of the wave functions the ANC is defined as the amplitude of the tail of the overlap function of the bound state wave functions of $B,\,A$ and $a$. The overlap function is given by 
\begin{align}
&I_{aA}^B({\rm {\bf r}}_{aA}) = \, <\psi_{c}\,|\,{\varphi _B}({\xi _A},\,{\xi _a};\,{\rm {\bf r}}_{aA})>       \nonumber\\
&= \,\sum\limits_{{l_B}{m_{{l_B}}}{j_B}{m_{{j_B}}}} { < {J_A}{M_A}\,\,{j_B}{m_{{j_B}}}|{J_B}{M_B} >  < {J_a}{M_a}\,{l_B}{m_{{l_B}}}|{j_B}{m_{{j_B}}} > } \,{Y_{{l_B}{m_{{l_B}}}}}({\rm {\bf {\widehat r}}}_{aA})\,I_{aA\,\,{l_B}{j_B}\,J_{B}}^B({r_{aA}}). 
\label{overlapfunction1}
\end{align} 
Here 
\begin{align}
\psi_{c}= \sum\limits_{m_{{j_B}}{m_{{l_B}}}{M_A}{M_{a}}} { < {J_A}{M_A}\,\,{j_B}{m_{{j_B}}}|{J_B}{M_B} >  < {J_a}{M_a}\,{l_B}{m_{{l_B}}}|{j_B}{m_{{j_B}}} > }\,{\widehat A_{aA}}\{ \varphi_{A}(\xi_{A})\,\varphi_{a}({\xi _a})\,Y_{ {l_B}\,{m_{l_{B}}} }({\rm {\bf {\widehat r}}}_{aA})\} 
\label{channelwf1}
\end{align}
is the two-body $a+ A$ channel wave function in the $jj$ coupling scheme, $<j_{1}\,m_{1}\,\,j_{2}\,m_{2}|j\,m>$ is the Clebsch-Gordan coefficient, ${\widehat A_{aA}}$ is the antisymmetrization operator between the nucleons of nuclei $a$  and $A$; $\,\,\varphi_{i}(\xi_{i})$ represents the fully antisymmetrized bound state wave function of nucleus $i$ with $\xi_{i}$ being a set of the internal coordinates including spin-isospin variables, $\,\,J_{i}$ and $M_{i}$ are spin and spin projection of nucleus $i$. Also ${\rm {\bf r}}_{aA}$ is the radius vector connecting the centers of mass of nuclei $a$ and $A$, $\,\,{\rm {\bf {\hat r}}}_{aA} = \,{\rm {\bf r}}_{aA}/r_{aA}$, $\,\,Y_{l_B\,m_{l_{B}}}({\rm {\bf {\hat r}}}_{Aa})$ is the spherical harmonics, and  $I_{aA\,\,{l_B}{j_B}J_{B}}^B({r_{Aa}})$ is the radial overlap function. The summation over ${l_B}$ and ${j_B}$ is carried out over the values allowed by the angular momentum and parity conservation in the virtual process $B \to A + a$.

The radial overlap function is given by
\begin{align}
&I_{aA\,\,{l_B}{j_B}\,J_{B}}^B({r_{aA}})\, = \, <{\hat A}_{aA}\,\{{\varphi _A}({\xi _A})\,{\varphi _a}({\xi_a})\,Y_{ {l_B}\,{m_{l_{B}}} }({\rm {\bf {\widehat r}}}_{aA})\}|\,{\varphi _B}({\xi _A},\,{\xi _a};\,{\rm {\bf r}}_{aA})>                                                 \nonumber\\
&= \left( \begin{gathered}
  A \hfill \\
  a \hfill \\ 
\end{gathered}  \right)^{\frac{1}{2}}<{\varphi _A}({\xi _A})\,{\varphi _a}({\xi_a})\,Y_{ {l_B}\,{m_{l_{B}}} }({\rm {\bf {\widehat r}}}_{aA})|\,{\varphi _B}({\xi _A},\,{\xi _a};\,{\rm {\bf r}}_{aA})>.
\label{radoverlap1}
\end{align}
Eq. (\ref{radoverlap1}) follows from a trivial observation that, because ${\varphi _B}$ is fully antisymmetrized, the antisymmetrization operator ${\hat A_{aA}}$ can be replaced by the factor  
$\left( \begin{gathered}
  A \hfill \\
  a \hfill \\ 
\end{gathered}  \right)^{\frac{1}{2}}$.
In what follows, in contrast to Blokhintsev et al (1977), we absorb this factor into the radial overlap function.     

The tail of the radial overlap function (${r_{aA}} > \,R_{aA}$) in the case of the normal asymptotic behavior is given by
\begin{align}
I_{aA\,\,{l_B}\,{j_B}\,J_{B}}^B({r_{aA}})\,\,\xrightarrow{{{r_{aA}} > {R_{aA}}}}\,\,{C^{B}_{aA\,{l_B}\,{j_B}\,J_{B}}}\frac{{{W_{ - {\eta _{aA}^{bs}},\,\,{l_B} + 1/2}}(2\,{\kappa_{aA}^{B}}{r_{aA}})}}{{{r_{aA}}}}\xrightarrow{{{r_{aA}} \to \infty }}{C^{B}_{aA\,{l_B}{j_B} J_{B}}}\,\frac{{{e^{ - \kappa _{aA}^B{r_{aA}} - {\kern 1pt} \eta _{aA}^{bs}\ln (2\kappa _{aA}^B{r_{aA}})}}}}{{{r_{aA}}}}.
\label{asymptoverlap1}
\end{align}
Correspondingly, for the resonance case
\begin{align}
&I_{aA\,\,{l_B}\,{j_B}\,J_{B}}^B({r_{aA}})\,\,\xrightarrow{{{r_{aA}} > {R_{aA}}}}\,\,{C^{B}_{aA\,{l_B}\,{j_B}\,J_{B}}}\frac{{{W_{ - i\,{\eta _{aA}^{R}},\,\,{l_B} + 1/2}}(-2\,i\,k_{aA\,(R)}{r_{aA}})}}{{{r_{aA}}}}  \nonumber\\
&\xrightarrow{{{r_{aA}} \to \infty }}{C^{B}_{aA\,{l_B}{j_B} J_{B}}}\,\frac{{{e^{ - i\,k_{aA\,(R)}\,{r_{aA}} - \,i\,\eta _{aA}^{R}\ln (-2\,i\,k_{aA\,(R)}\,{r_{aA}})}}}}{{{r_{aA}}}}.
\label{asymptoverlapres1}
\end{align}

For the first time, the proof of the fundamental connection between the residue $A_{l_{B}\,j_{B}}^{J_{B}}$ in the bound state pole of the elastic scattering $S$ matrix and the amplitude $C^{B}_{aA\,{l_B}\,{j_B}\,J_{B}}$ of the tail of the overlap function has been presented in \cite{kramers} and later on in \cite{heisenberg,meller,hu} \footnote[1]{As a matter of fact, in these works the ANC was the amplitude of the bound state wave function of two structureless particles.}.
The proof of this relationship for non-sperical potentials was given in \cite{zeldovich} and for charged particles in \cite{perelomov}. Finally, generalization of this relationship for general case of bound state and resonances 
for charged particles was presented in \cite{muk77}.

The first comprehensive review about the overlap functions and ANC was given in Ref. \cite{blokh77}, in which the theory of the ANC and its role in the nuclear reaction theory was presented. In another review paper \cite{blokhintsev84} the role of the ANC in the theory of nuclear reactions with charged particles was addressed.
The role of the ANC in nuclear astrophysics was, for the first time, discussed in \cite{muktim90,muk90}, where it was underscored that the ANC determines the overall normalization of the peripheral radiative capture reactions. These two works pawed the way for using the ANC method as indirect method in nuclear astrophysics (see also \cite{hu94}). The ANC can be determined from peripheral transfer reactions and can be used to calculate peripheral radiative capture reactions. It constitute 
a powerful indirect ANC method in nuclear astrophysics. The ANC method was extensively used in the analysis of many important astrophysical reactions (see, for example, \cite {muk95,tang,muk2003,muk2006,banu,abdullah,muk08,muk2011,kroha2011} and references therein). 
The role of the ANC in nuclear astrophysics has been once again underscored in the review \cite{adelberger}.

Recently it was shown that in exact approach the only model-independent quantity, which can be determined from the analysis of nuclear reactions, is the ANC rather than the spectroscopic factor  \cite{muk2010}. 
Moreover, in \cite{mukstripping2011} a new formalism of the deuteron stripping based on the surface integral formalism and generalized $R$-matrix approach has been developed. It makes the role of the ANC in nuclear reaction theory even more important than it was thought before. 
In this formalism the amplitude of the deuteron stripping reactions populating bound states and resonances is paremeterized in terms of the ANC rather than the spectroscopic factors.

In \cite{timofeyuk2003} the charge symmetry of strong interactions was used to relate the proton and neutron ANCs of the one-nucleon overlap integrals for light mirror nuclei. Equation (7) from \cite{timofeyuk2003}, determining the ratio of the proton and neutron ANCs of mirror nuclei, allows us to find one of the ANCs if another one is known. This relation extends to the case of real proton decay where the mirror analog is a virtual neutron decay of a loosely bound state. In this case, a link is obtained between the proton width and the squared ANC of the mirror neutron state. The relation between mirror overlaps was used to study astrophysically relevant proton capture reactions based on information obtained from transfer reactions with stable beams \cite{tim2005,tim2006}. 

In a nice work \cite{nazarewicz} the impact of the particle continuum on the proton and neutron ANCs for mirror $p$- and $d$-shell nuclei within the framework of the real-energy and complex- energy continuum shell-model approaches. The authors consider the basic properties of the single-particle ANCs for charged and neutral particles as functions of the Coulomb parameter,
binding energy and the orbital angular momentum. The authors investigate the validity of the ratio of the proton and neutron mirror ANCs given by Eq. (7) from \cite{timofeyuk2003}. The main finding of \cite{nazarewicz} is that the key factor affecting the ratio of the mirror ANCs is the distribution of the spectroscopic strength.   

In Ref. \cite{titus} the model independence of the ratio of the proton and neutron ANCs for mirror nuclei was tested  within a phenomenological model, where the valence nucleon moves in a deformed mean field of the core which is allowed to excite.

The calculations show that the ratio of the proton and neutron ANCs (see Table II from \cite{timofeyuk2003}) increases dramatically with decrease of the proton binding energy and the charge of the core. 
This is the result of the impact of the Coulomb barrier, which blocks the proton bound state wave function in the nuclear interior significantly decreasing its tail. Because the wave function is still normalized to unity, the drop of the radial tail is compensated by corresponding increase of the amplitude of the tail. For loosely bound states and high charges the conventionally determined ANC becomes enormously huge making problems in computations. For example the square of the proton ANC  ${}^{21}{\rm Na}(1/2^{+},\,2.425 {\rm MeV}) \to {}^{20}{\rm Ne} + p$ is $6.14 \times 10^{33}$ fm${}^{-1}$ \cite{muk2006}. The square of the ANC for ${}^{17}{\rm O }(6.356\,{\rm MeV},\,1/2^{+}) \to {}^{13}{\rm C} + \alpha$ is $\sim 10^{168}$ fm${}^{-1}$ \cite{johnson2006}. 
In this paper I will give a physical insight into factors determining the Coulomb renormalization of the ANC,  making easier comparison of the ANCs of mirror nuclei. The analysis of the Coulomb renormalization of the ANC is done using two approaches: the model-independent analytic structure of the scattering amplitude and the Pinkston-Satchler equation  \cite{pinkston,philpott}. All the factors, which affect the ANC owing to the Coulomb interaction, are analyzed. As a result of this research I suggest to use in the future analysis of the data the renormalized proton ANC, in which the main Coulomb renormalization factor is excluded. It is especially useful in cases of high nuclei charges and low binding energies. I provide also a new expression for the ratio of the mirror ANCs based on the transformation of the expression for the overlap function obtained from the Pinkston-Satchler equation in terms of the Wronskian. 
This expression is especially useful when microscopic overlap functions are available. These microscopic overlap functions are usually accurate in the nuclear interior what is enough to determine the ratio of the ANCs of the mirror nuclei expressed in terms of the Wronskian. Then if one of the ANCs is known experimentally the second one can be determined. 

\section{Coulomb renormalization of the ANC from analytic structure of the scattering amplitude}
\label{analyticstructure1}

First, I address the Coulomb renormalization of the ANC from analytic structure of the scattering amplitude \cite{blokhintsev84}. Let us consider the scattering of two charged spinless particles $a$ and $A$ (we disregard the spins because they don't affect the Coulomb renormalization). The analytic properties of the on-the-energy-shell Coulomb-nuclear partial wave scattering amplitude was investigated in \cite{wong,mentkovsky,hamilton,baz,blokhintsev83}. The elastic scattering $S$-matrix element is given by
\begin{align}
S_{l_{B}}(k_{aA})= 1 + 2\,i\,\rho_{l_{B}}(k_{aA})\,F_{l_{B}}(k_{aA}),
\label{Smatrix1}
\end{align}
where $\rho_{l_{B}}(k_{aA})= k_{aA}^{2\,l_{B}+1}$.
The partial elastic scattering $S$-matrix element can be written as 
\begin{align}
S_{l_{B}}(k_{aA})= e^{2\,i\,\delta_{l_{B}}}= S_{l_{B}}^{C}(k_{aA}) + S_{l_{B}}^{C}(k_{aA})(S_{l_{B}}^{CN}(k_{aA}) -1),
\label{Slk1}
\end{align}
\begin{align}
S_{l_{B}}^{C}(k_{aA})= e^{2\,i\,\delta_{l_{B}}^{C}}= \frac{\Gamma(l_{B}+ 1 + i\,\eta_{aA})}{\Gamma(l_{B}+ 1 - i\,\eta_{aA})}
\label{CoulombSmatrel1}
\end{align}
is the Coulomb elastic scattering $S$-matrix element and 
\begin{align}
S_{l_{B}}^{CN}(k_{aA})= e^{2\,i\,\delta_{l_{B}}^{CN}}
\label{SlCN1}
\end{align}
is the Coulomb-modified nuclear elastic scattering $S$-matrix element, $\,\eta_{aA}= Z_{1}\,Z_{2}\,e^{2}\,\mu_{aA}/k_{aA}\,$ is the Coulomb parameter of particles $a$ and $A$ in continuum, 
$\,\delta_{l_{B}}$ is the total scattering phase shift in the partial wave $l_{B}$, and $\delta_{l_{B}}^{CN}= \delta_{l_{B}} - \delta_{l_{B}}^{C}$ is the Coulomb-modified nuclear scattering phase shift.
Coulomb-modified elastic scattering $S$-matrix element $S_{l_{B}}^{CN}(k_{aA})$ we rewrite as  
\begin{align}
S_{l_{B}}^{CN}(k_{aA})= e^{2\,i\,\delta_{l_{B}}}= 1 + 2\,i\,\rho_{l_{B}}(k_{aA})\,F_{l_{B}}^{CN}(k_{aA}).
\label{smatrixel1}
\end{align}
The Coulomb-modified nuclear partial wave scattering amplitude $F_{l_{B}}^{CN}(k_{aA})$
is not analytical function in the $k_{aA}$ plane \cite{mentkovsky}.
Following \cite{mentkovsky,blokhintsev84} we can present  
\begin{align}
F_{l_{B}}^{CN}(k_{aA})= \frac{1}{[h_{l_{B}}^{C}(k_{aA})]^{2}}\,T_{l_{B}}^{CN}(k_{aA}).
\label{CNelstampl1}
\end{align} 
Here
\begin{align}
h_{l_{B}}^{C}(k_{aA})= \frac{l_{B}!\,e^{\frac{\pi}{2}\,\eta_{aA}\,{\rm sign}\,[Re\,k_{aA}]}}{\Gamma(l_{B}+ 1 + i\,\eta_{aA})}
\label{hlck1}
\end{align}
is the normalized ($h_{l_{B}}^{C}(k_{aA}) \xrightarrow{k_{aA} \to \infty} 1$) Coulomb Jost function.
At real $k_{aA} >0$ this definition coincides with a conventional definition of the Coulomb Jost function.
Owing to the presence of the factor ${\rm sign}\,[Re\,k_{aA}]$, $F_{l_{B}}^{CN}(k_{aA})$ has singularity along the imaginary axis in the $k_{aA}$-plane, which splits $F_{l_{B}}^{CN}(k_{aA})$ into two parts corresponding to ${\rm Re}\,k_{aA} >0$ and ${\rm Re}\,k_{aA}<0$.   

However the singled out renormalized Coulomb-modified nuclear partial wave scattering amplitude $T_{l_{B}}^{CN}(k_{aA})$ has the same analytical properties on the first (physical) sheet of the Reeman surface ($Im\,k>0$) as the pure nuclear partial wave scattering amplitude:
right-hand unitarity cut ($0 \leq k_{aA} < \infty$) and left-hand ($E_{aA} <0$) dynamical cut \cite{blokhintsev84}.
$T_{l_{B}}^{CN}(k_{aA})$, similar to the pure nuclear scattering amplitude, has a pole at $k_{aA}=i\,\kappa_{aA}^{B}$ 
\begin{align}
T_{l_{B}}^{CN}(k_{aA}) \xrightarrow{k_{aA} \to i\,\kappa_{aA}^{B}} \frac{{\tilde A}_{l_{B}}}{k_{aA} - i\,\kappa_{aA}^{B}},
\label{TlCNpole1}
\end{align}
where the residue in the pole
\begin{align}
{\tilde A}_{l_{B}}= -i^{2\,l_{B}+1}\,\big({\tilde C}^{B}_{aA\,l_{B}}\big)^{2}.
\label{residuepole1}
\end{align}
Correspondingly, the residue of $F_{l_{B}}^{CN}(k_{aA})$ is 
\begin{align}
A_{l_{B}}= -i^{2\,l_{B}+1}\,e^{i\,\pi\,\eta_{aA}^{bs}}\,\big[\frac{\Gamma(l_{B}+ 1 + \eta_{aA}^{bs})}{l_{B}!}\big]^{2}\,\big({\tilde C}^{B}_{aA\,l_{B}}\big)^{2}.
\label{residue2}
\end{align}
Here ${\tilde C}^{B}_{aA\,l_{B}}$ is the renormalized ANC, which is connected, according to Eq. (\ref{hlck1}), to the standard ANC \cite{blokhintsev84}:
\begin{align}
C^{B}_{aA\,l_{B}}= \frac{\Gamma(l_{B} + 1 + \eta_{aA}^{bs})}{l_{B}!}\,{\tilde C}^{B}_{aA\,l_{B}}.
\label{standardANC11}
\end{align}
As we can see, the standard ANC contains the Coulomb barrier factor $\frac{\Gamma(l_{B} + 1 + \eta_{aA}^{bs})}{l_{B}!}$. Correspondingly,
\begin{align}
\big(C^{B}_{aA\,l_{B}}\big)^{2}= {\Rcu}_{1}\,\big({\tilde C}^{B}_{aA\,l_{B}}\big)^{2} .
\label{standardANC1}
\end{align}
where the renormalization factor in Eq. (\ref{standardANC1}) is
\begin{align}
{\Rcu}_{1}= \Big[\frac{\Gamma(l_{B}+ 1 + \eta_{aA}^{bs})}{l_{B}!} \Big]^{2},
\label{CRF1}
\end{align}
which is derived only from consideration of the analytic properties of the elastic scattering amplitude. This factor is the main Coulomb renormalization factor (CRF) of the conventional ANC due to the Coulomb barrier. It increases with increasing of charges and decreasing of the binding energy and can be quite huge. 

Now can rewrite the reduced width as (recovering the spins)
\begin{align}
\gamma _{aA\,{l_B}\,{j_B}\,J_{B}}^2 = \frac{{{\tilde W}_{ - {\eta _{aA}^{bs}},\,{l_B}\, + \,1/2}^2(2{\kappa _{aA}^{B}}{R_{aA}})\,}}{{2{\mu _{aA}}{R_{aA}}}}\,\big({\tilde C}^{B}_{aA\,{l_B}\,{j_B}\,J_{B}}\big)^2,
\label{redwidth1}
\end{align}
where 
\begin{align}
{\tilde W}_{ - {\eta _{aA}^{bs}},\,{l_B}\, + \,1/2}^2(2{\kappa _{aA}^{B}}{R_{aA}}) =
\frac{\Gamma(l+ 1 + \eta_{aA}^{bs})}{l!}\,W_{ - {\eta _{aA}^{bs}},\,{l_B}\, + \,1/2}^2(2{\kappa _{aA}^{B}}{R_{aA}}).
\label{tildeW1}
\end{align}
The reduced width is given by the product of two factors. When the charge of nucleus increases and binding energy 
$\varepsilon_{aA}^{B}$ decreases, the Coulomb barrier also increases decreasing significantly the tail of the Whittaker function. However, owing the conservation of the normalization of the bound state wave function of nucleus $B$, the amplitude of the tail of the overlap function (its ANC) also significantly increases. Renormalization of the ANC and the Whittaker function, as it is done in Eq. (\ref{tildeW1}), decreases the ANC
and increases the tail of the Whittaker function. This renormalization doesn't change the reduced width but allows one to operate with smaller ANCs then the standard ones and I draw attention of the experimentalists on the possibility to use the Coulomb renormalized ANCs rather than the standard ones when the latter become too big.

I demonstrate the renormalization of the ANC in Table \ref{table_chiS70}. 
\begin{table}
\begin{center}
\caption{The squared proton ANCs $\big(C^{B}_{pA\,l_{B}j_{B}J_{B}}\big)^{2}$ and Coulomb renormalized ANCs $\big({\tilde C}^{B}_{pA\,l_{B}j_{B}J_{B}}\big)^{2}$ of the overlap functions $I_{pA\,l_{B}j_{B}J_{B}}^{B}$; $\,(l_{B}j_{B}J_{B})$ are the quantum numbers of the removed proton and the total spin of nucleus $B$; $\varepsilon_{pA}^{B}$ is the binding energy for the virtual decay $B \to p + A$. The ANCs are taken from \cite{tim2011}.  
\label{table_chiS70}} 
\begin{tabular}{ccccccc}
\hline \hline \\
 B & \,\,A\, \,&\,$l_{B}j_{B}$ & $J_{B}$  &    $\varepsilon_{pA}^{B}$ (MeV)  &  $\big(C^{B}_{pA\,l_{B}j_{B}}\big)^{2}$ fm${}^{-1}$ & $\big({\tilde C}^{B}_{pA\,l_{B}j_{B}}\big)^{2}$ fm${}^{-1}$        \\
\hline\\[1mm]\\ 
${}^{21}{\rm Na}$\,\,&\,${}^{20}{\rm Ne}(0.0\,{\rm MeV})$\,\,&\,\, $s_{1/2}$ &  $1/2$ &   $0.0071$  \,\, & $6.5 \times 10^{33}$  &$2.66$ \\
\hline\\[1mm]\\ 
${}^{57}{\rm Cu}$\,\,&\,${}^{56}{\rm Ni}(0.0\,{\rm MeV})$\,\,&\,\, $p_{3/2}$ &  $3/2$ & $0.7$  \,\, & $1.77 \times 10^{8}$  & $135$ \\
\hline\\[1mm]\\ 
${}^{132}{\rm Sn}$\,\,&\,${}^{131}{\rm In}(0.0\,{\rm MeV})$\,\,&\,\, $g_{9/2}$ & $0$  & $15.71$ \,\, & $9.03 \times 10^{8}$  & $1.22 \times 10^{6}$ \\
[2mm]
\hline \hline \\
\end{tabular}
\end{center}
\end{table}

\section{ANC from Pinkston-Satchler equation}
\label{ANCsourceequation1}

\subsection{Coulomb renormalization of ANC from source term expression}

Here we obtain the main CRF of the ANC using the equation for the ANC containing 
the source term \cite{muk90,tim1998}. To make it more clear, first we consider the derivation of this expression following \cite{muk90,tim1998} from Pinkston-Satchler equation \cite{pinkston,philpott}. Note that the overlap function is not an eigenfunction of any Hermitian Hamiltonian. To derive equation for the overlap function containing the source term we start from the Schr\"odinger equation for the bound state of the parent nucleus $A$:
\begin{align}
({E_B} - {\widehat T_A} - {\widehat T_a} - {\widehat T_{aA}} - \,V_{a}  - \,V_{A} - V_{aA} ){\varphi _B}({\xi _A},{\xi _a};\,{\rm {\bf r}}_{aA}) = 0.                             
\label{Schreq1}              .                  
\end{align}
Here, ${\widehat T_i}$ is the internal motion kinetic energy operator of nucleus 
$i$, $\,{\widehat T_{aA}}$ is the kinetic energy operator of the relative motion of nuclei 
$a$ and $A$, $V_{i}$ is the internal potential of nucleus $i$ and $V_{aA}$ is the interaction potential between $a$ and $A$, $\,\,E_{B}$ is the total binding energy of nucleus $B$.

Multiplying the Schr\"odinger equation (\ref{Schreq1})  from the left by  
\begin{align}
\,{\left( \begin{gathered}
  A \hfill \\
  a \hfill \\ 
\end{gathered}  \right)^{1/2}}\,\sum\limits_{{m_{{j_B}}}{m_{{l_B}}}M_{A}M_{a}} { < {J_A}{M_A}\,\,{j_B}{m_{{j_B}}}|{J_B}{M_B} >  < {J_a}{M_a}\,{l_B}{m_{{l_B}}}|{j_B}{m_{{j_B}}} > } \,{Y^{*}_{{l_B}{m_{{l_B}}}}}({\rm {\bf {\widehat r}}}_{aA})\,\varphi_{A}(\xi_{A})\,\varphi_{a}(\xi_{a})
\label{projectaAstate1}
\end{align}
and taking into account Eq. (\ref{radoverlap1})
we get the equation for the radial overlap function with the source term \cite{tim1998}
\begin{align}
\Big(\varepsilon_{aA}^{B} - {\hat T}_{r_{aA}}- V^{centr}_{l_{B}} - U_{aA}^{C} \Big)\,I_{aA\,\,{l_B}\,{j_B}\,J_{B}}^{B}(r_{aA}) = Q_{l_{B}j_{B}J_{a}J_{A}J_{B}}(r_{aA}).
\label{radeqst1}
\end{align}
Here, ${\hat T}_{r_{aA}}$ is the radial relative kinetic energy operator of the particles $a$ and $A$, $\,V_{l_{B}}^{centr}$ is the centrifugal barrier for the relative motion of $a$ and $A$ with the orbital momentum $l_{B}$;  $\,Q_{l_{B}j_{B}J_{a}J_{A}J_{B}}(r_{aA})$ is the source term
\begin{align}
&Q_{l_{B}j_{B}J_{a}J_{A}J_{B}}(r_{aA}) = \sum\limits_{m_{j_B}m_{l_B}M_{A}M_{a}} { < {J_A}{M_A}\,\,{j_B}{m_{{j_B}}}|{J_B}{M_B} >  < {J_a}{M_a}\,{l_B}{m_{{l_B}}}|{j_B}{m_{{j_B}}} > }\nonumber\\
& \times\,{\left( \begin{gathered}A \hfill \\
a \hfill \\ 
\end{gathered}  \right)^{1/2}}\,\int\,{\rm d}\,\Omega_{{\rm {\bf r}}_{aA}}\,<\varphi_{a}(\xi_{a})\,\varphi_{A}(\xi_{A})|V_{aA} - U_{aA}^{C}|Y_{{l_B}{m_{{l_B}}}}^{*}({\rm {\bf {\widehat r}}}_{aA})\varphi_{B}(\xi_{a},\xi_{A};{\rm {\bf r}}_{aA})>.
\label{sourceeqn1}
\end{align}
The integration in the matrix element $<\varphi_{a}(\xi_{a})\,\varphi_{A}(\xi_{A})|V_{aA} - U_{aA}^{C}|Y_{{l_B}{m_{{l_B}}}}^{*}({\rm {\bf {\widehat r}}}_{aA})\varphi_{B}(\xi_{a},\xi_{A};{\rm {\bf r}}_{aA})>\,\,$ in Eq. (\ref{sourceeqn1}) is carried out over all the internal coordinates of nuclei $a$ and $A$.  Note that we replaced the antisymmetrization operator ${\hat A}_{aA}$ in Eq. (\ref{projectaAstate1}) by  ${\left( \begin{gathered}A \hfill \\
a \hfill \\ 
\end{gathered}  \right)^{1/2}}$ because the operator ${E_B} - {\widehat T_A} - {\widehat T_a} - {\widehat T_{aA}} - \,V_{a}  - \,V_{A} - V_{aA}$ in Eq. (\ref{Schreq1}) is symmetric over interchange of nucleons of $a$ and $A$, while $\varphi_{B}$ is antisymmetric. 
For charged particles it is convenient to single out the channel Coulomb interaction $U_{aA}^{C}(r_{aA})$ between the center of mass of nuclei $a$ and $A$.
 
Eq. (\ref{sourceeqn1}) can be rewritten in the integral form:
\begin{align}
I_{aA\,\,{l_B}\,{j_B}\,J_{B}}^{B}(r_{aA}) = \,\frac{1}{r_{aA}}\,\int\limits_{0}^{\infty}\,{\rm d}r'_{aA}\,r'_{aA}\,G_{l_{B}}^{C}(r_{aA},\,r'_{aA}; -\varepsilon_{aA}^{B})\,Q_{l_{B}j_{B}J_{a}J_{A}J_{B}}(r'_{aA}).
\label{radeqst2}
\end{align}
The partial Coulomb two-body Green function is given by  \cite{newton}
\begin{align}
G_{l_{B}}^{C}(r_{aA},r'_{aA}; -\varepsilon_{aA}^{B})= -2\,\mu_{aA}\,\frac
{\varphi_{l_{B}}^{C}(i\,\kappa_{aA}^{B}\,r_{aA\,<})\,f_{l_{B}}^{C(+)}(i\,\kappa_{aA}^{B}\,r_{aA\,>})}{L_{l_{B}}^{C(+)}},
\label{Gredenfunct1}
\end{align}
where $\,\,r_{aA\,<} = {\rm  min}\,\{r_{aA},r'_{aA}\}$ and $\,\,r_{aA\,>}= {\rm max}\,\{r_{aA},r'_{aA} \}$. The Coulomb regular solution $\varphi_{l_{B}}^{C}(i\,\kappa_{aA}^{B}\,r_{aA})$ of the partial Schr\"odinger equation at imaginary momentum $i\,\kappa_{aA}^{B}$ is
\begin{align}
&\varphi_{l_{B}}^{C}(i\,\kappa_{aA}^{B}\,r_{aA}) = - \frac{1}{2\,\kappa_{aA}^{B}}\,\Big[L_{l_{B}}^{C(-)}(i\,\kappa_{aA}^{B})\,f_{l_{B}}^{C(+)}(i\,\kappa_{aA}^{B},r_{aA}) - L_{l_{B}}^{C(+)}(i\,\kappa_{aA}^{B})\,f_{l_{B}}^{C(-)}(i\,\kappa_{aA}^{B}\,r_{aA})\Big]                                                          \nonumber\\
&= r_{aA}^{l_{B}+1}\,e^{-\,\kappa_{aA}^{B}\,r_{aA}}\,{}_1F_{1}(l_{B} + 1 + \eta_{aA}^{bs}, 2\,l_{B} + 2; 2\,\kappa_{aA}^{B}\,r_{aA})                                    \nonumber\\
&= e^{-i\,\pi\,l_{B}/2}\,L_{l_{B}}^{C(+)}(i\,\kappa_{aA}^{B})\,\frac{e^{i\,\sigma_{l_{B}}}\,F_{l_{B}}(i\,\kappa_{aA}^{B},r_{aA})}{i\,\kappa_{aA}^{B}},                                                 
\label{varphiFl1}
\end{align}
where
\begin{align}
e^{i\,\sigma_{l_{B}}}\,F_{l_{B}}(i\,\kappa_{aA}^{B},r_{aA})=e^{i\,\pi\,\eta_{aA}^{bs}/2}\,\frac{\Gamma(l_{B} + 1 + \,\eta_{aA}^{bs})}{2\,\Gamma(2\,l_{B} + 2)}\,(2\,i\,\kappa_{aA}^{B}\,r_{aA})^{l_{B} + 1}\,e^{-\,\kappa_{aA}^{B}\,r_{aA}}\,{}_1F_{1}(l_{B} + 1 + \,\eta_{aA}^{bs}, 2\,l_{B} + 2; \, 2\,\kappa_{aA}^{B}\,r_{aA}).
\label{Coulregfunct1}
\end{align}
Also
\begin{align}
f_{l_{B}}^{C(\pm)}(i\,\kappa_{aA}^{B},r_{aA}) = e^{-i\,\pi\,\eta_{aA}^{bs}/2}\,W_{\mp \eta_{aA}^{bs},\,l_{B} + 1/2}(\pm 2\,\kappa_{aA}^{B}\,r_{aA})
\label{iostsolution1}
\end{align}
are the Jost solutions (singular at the origin $r_{aA}=0$), 
\begin{align}
L_{l_{B}}^{C(\pm)}(i\,\kappa_{aA}^{B}) = \frac{1}{(2\,i\,\kappa_{aA}^{B})^{l_{B}}}\,e^{-i\,\pi\,\eta_{aA}^{bs}/2}\,e^{\pm i\,\pi\,l_{B}/2}\,\frac{\Gamma(2\,l_{B} + 2)}{\Gamma(l_{B} +1 \pm \eta_{aA}^{bs})}
\label{iostfunction1}
\end{align}
are the Jost functions.

The asymptotic behavior of the overlap function is correct because it is governed by the Green function:
\begin{align}
I_{aA\,\,{l_B}\,{j_B}\,J_{B}}^{B}(r_{aA}) \stackrel{r_{aA} \to \infty}{\approx} \,-\,2\,\mu_{aA}\,\frac{W_{-\eta_{aA}^{B},\,l_{B} + 1/2}(2\,\kappa_{aA}^{bs}\,r_{aA})}{r_{aA}}\,\frac{ e^{-i\,\pi\,\eta_{aA}^{bs}/2}}{L_{l_{B}}^{C(+)}(i\,\kappa_{aA}^{B})}\,\int\limits_{0}^{r_{aA}}\,{\rm d}r'_{aA}\,r'_{aA}\,\varphi_{l_{B}}^{C}(i\,\kappa_{aA}^{B}\,r_{aA}')\,Q_{l_{B}j_{B}J_{a}J_{A}J_{B}}(r'_{aA}).
\label{asymradovint1}
\end{align}

Taking into account Eq. (\ref{asymptoverlap1}) we get the ANC expressed in terms of the source $Q_{l_{B}j_{B}J_{a}J_{A}J_{B}}(r_{aA})$ \cite{muk90,tim1998}:
\begin{align}
C_{aA\,\,{l_B}\,{j_B}\,J_{B}}^{B} = -\,2\,\mu_{aA}\,e^{-i\,\pi\,\eta_{aA}^{bs}/2}\,\frac{1}{L_{l_{B}}^{C(+)}(i\,\kappa_{aA}^{B})}\int\limits_{0}^{\infty}\,{\rm d}r_{aA}\,r_{aA}\,\varphi_{l_{B}}^{C}(i\,\kappa_{aA}^{B}\,r_{aA})\,Q_{l_{B}j_{B}J_{a}J_{A}J_{B}}(r_{aA}).
\label{ANCst1}
\end{align}
Owe to the presence of the short-range potential operator $V_{aA} - U_{aA}^{C}$ (potential $V_{aA}$ is the sum of the nuclear $V^{N}_{aA}$ and the Coulomb $V_{aA}^{C}$ potentials and subtraction of $U_{aA}^{C}$ removes the long-range Coulomb term from $V_{aA}$) the source term is also a short-range function and we approximate Eq. (\ref{ANCst1}) by 
\begin{align}
C_{aA\,\,{l_B}\,{j_B}\,J_{B}}^{B} \approx -\,2\,\mu_{aA}\,e^{-i\,\pi\,\eta_{aA}^{bs}/2}\,\frac{1}{L_{l_{B}}^{C(+)}(i\,\kappa_{aA}^{B})}\,\int\limits_{0}^{R_{aA}}\,{\rm d}r_{aA}\,r_{aA}\,\varphi_{l_{B}}^{C}(i\,\kappa_{aA}^{B}\,r_{aA})\,Q_{l_{B}j_{B}J_{a}J_{A}J_{B}}(r_{aA}),
\label{ANCstapprox1}
\end{align}
where $R_{aA}$ is the channel radius. 
This equation provides the ANC, which, as it will be shown below, depends on $R_{aA}$ and may not be accurate enough because we cut the integration over $r_{aA}$ at the channel radius $R_{aA}$. However, here we are interested in the ratio of the mirror proton and neutron ANCs and, as it will be demonstrated below, this ratio is practically insensitive to the value of the channel radius.

$\big[L_{l_{B}}^{C(+)}(i\,\kappa_{aA}^{B})\big]^{-1}$  contains the same barrier 
$\Gamma(l_{B} + 1 + \eta_{aA}^{bs})$, see Eq. (\ref{iostfunction1}), which has been 
found in Section \ref{analyticstructure1} based on the general principle of analiticity of the elastic scattering amplitude. Hence, following Eq. (\ref{standardANC11}), we introduce 
the renormalized ANC
\begin{align}
{\tilde C}_{aA\,\,{l_B}\,{j_B}\,J_{B}}^{B} \approx -\,2\,\mu_{aA}\,(2\,\kappa_{aA}^{B})^{l_{B}}\,\,\frac{\Gamma(l_{B} +1)}{\Gamma(2\,l_{B} + 2)}\,\int\limits_{0}^{R_{aA}}\,{\rm d}r_{aA}\,r_{aA}\,\varphi_{l_{B}}^{C}(i\,\kappa_{aA}^{B}\,r_{aA})\,Q_{l_{B}j_{B}J_{a}J_{A}J_{B}}(r_{aA}).
\label{tildeANC1}
\end{align}
The conventional ANC is related to the renormalized one as  
\begin{align}
C_{aA\,\,{l_B}\,{j_B}\,J_{B}}^{B}= \frac{\Gamma(l_{B} + 1 + \eta_{aA}^{bs})}{l_{B}!}\,{\tilde C}_{aA\,\,{l_B}\,{j_B}\,J_{B}}^{B}.
\label{standANCtildeANC1}
\end{align}

\subsection{ANC in terms of Wronskian}
\label{wronskian}

The advantage of Eq. (\ref{ANCstapprox1}) is that to calculate the ANC one needs to know the overlap function only in the nuclear interior where the {\it ab initio} methods like no-core-shell-model \cite{navratil2000,navratil2003,quaglioni}, variational Green function Monte Carlo method \cite{pieper,nollett2007,nollett2011} and coupled-cluster method \cite{jensen}  are more accurate than in the external region. Now we transform the radial integral in Eq. (\ref{ANCstapprox1}) into the Wronskian at $r_{aA}= R_{aA}$. The philosophy of this transformation is the same as in the surface integral formalism \cite{kadyrov2009,mukkadyrov,strippingresonance2011} only applied here for bound states.

First we rewrite
\begin{align}
V_{aA}- U_{aA}^{C}= V + V_{l_{B}}^{centr}- V_{a} - V_{A} - V_{l_{B}}^{centr} - U_{aA}^{C}
\label{pottransf1}
\end{align} 
and take into account equations 
\begin{align}
(-\varepsilon_{B} - {\hat T}_{a} - {\hat T}_{A} - {\hat T}_{r_{aA}})\,\varphi_{l_{B}}^{C}(i\,\kappa_{aA}^{B}\,r_{aA})\,\varphi_{a}(\xi_{a})\,\varphi_{A}(\xi_{A})= (U_{aA}^{C} + V_{l_{B}}^{centr} + V_{a} + V_{A})\,\varphi_{l_{B}}^{C}(i\,\kappa_{aA}^{B}\,r_{aA})\,\varphi_{a}(\xi_{a})\,\varphi_{A}(\xi_{A})
\label{Shreq2}
\end{align}
and 
\begin{align}
(-\varepsilon_{B} - {\hat T}_{a} - {\hat T}_{A} - {\hat T}_{r_{aA}})\,<Y_{{l_B}{m_{{l_B}}}}({\rm {\bf {\widehat r}}}_{aA})|\varphi_{B}>\,=\,(V_{aA} + V_{a} + V_{A} + V_{l_{B}}^{centr})\,<Y_{{l_B}{m_{{l_B}}}}({\rm {\bf {\widehat r}}}_{aA})|\varphi_{B}>,  
\label{eqvarphiB1}
\end{align}
where ${\hat T}_{r_{aA}}$ is the radial kinetic energy operator.
Then we get 
\begin{align}
&C_{aA\,\,{l_B}\,{j_B}\,J_{B}}^{B} \approx -\,2\,\mu_{aA}\,e^{-i\,\pi\,\eta_{aA}^{bs}/2}\,\frac{1}{L_{l_{B}}^{C(+)}(i\,\kappa_{aA}^{B})}\,\int\limits_{0}^{R_{aA}}\,{\rm d}r_{aA}\,r_{aA}\,               
\varphi_{l_{B}}^{C}(i\,\kappa_{aA}^{B}\,r_{aA})\,Q_{l_{B}j_{B}J_{a}J_
{A}J_{B}}(r_{aA})\,=  \,-\,2\,\mu_{aA}\,e^{-i\,\pi\,\eta_{aA}^{bs}/2}                                                          \nonumber\\
& \times\,\frac{1}{L_{l_{B}}^{C(+)}(i\,\kappa_{aA}^{B})}\,\sum\limits_{{m_{j_B}{m_{{l_B}}}M_{A}M_{a}}} { < {J_A}{M_A}\,\,{j_B}{m_{{j_B}}}|{J_B}{M_B} >  < {J_a}{M_a}\,{l_B}{m_{{l_B}}}|{j_B}{m_{{j_B}}} > }\,{\left( \begin{gathered}A \hfill \\
a \hfill \\ 
\end{gathered}  \right)^{1/2}}\,\int\limits_{0}^{R_{aA}}\,{\rm d}r_{aA}\,r_{aA}\,\varphi_{l_{B}}^{C}(i\,\kappa_{aA}^{B}\,r_{aA})                  \nonumber\\
&\times\,\int\,{\rm d}\,\Omega_{{\rm {\bf r}}_{r_{aA}}} \,<\varphi_{a}(\xi_{a})\,\varphi_{A}(\xi_{A})|{\overleftarrow {\hat T}}_{r_{aA}} + {\overleftarrow {\hat T}}_{a} + {\overleftarrow {\hat T}}_{A} - {\overrightarrow {\hat T}}_{a} - {\overrightarrow {\hat T}}_{A} - {\overrightarrow {\hat T}}_{r_{aA}}|Y_{{l_B}{m_{{l_B}}}}^{*}({\rm {\bf {\widehat r}}}_{aA})\,\varphi_{B}(\xi_{a},\xi_{A};{\rm {\bf r}}_{aA})>                                                                       \nonumber\\
&= -\,2\,\mu_{aA}\,e^{-i\,\pi\,\eta_{aA}^{bs}/2}\,\frac{1}{L_{l_{B}}^{C(+)}(i\,\kappa_{aA}^{B})}\,\sum\limits_{{m_{j_B}{m_{{l_B}}}M_{A}M_{a}}} { < {J_A}{M_A}\,\,{j_B}{m_{{j_B}}}|{J_B}{M_B} >  < {J_a}{M_a}\,{l_B}{m_{{l_B}}}|{j_B}{m_{{j_B}}} > }\nonumber\\
& \times\,{\left( \begin{gathered}A \hfill \\
a \hfill \\ 
\end{gathered}  \right)^{1/2}}\,\int\limits_{0}^{R_{aA}}\,{\rm d}r_{aA}\,r_{aA}\,\varphi_{l_{B}}^{C}(i\,\kappa_{aA}^{B}\,r_{aA})\,\int\,{\rm d}\,\Omega_{{\rm {\bf r}}_{aA}}\,<\varphi_{a}(\xi_{a})\,\varphi_{A}(\xi_{A})|{\overleftarrow {\hat T}}_{r_{aA}} - {\overrightarrow {\hat T}}_{r_{aA}}|Y_{{l_B}{m_{{l_B}}}}^{*}({\rm {\bf {\widehat r}}}_{aA})\varphi_{B}(\xi_{a},\xi_{A};{\rm {\bf r}}_{aA})>
\nonumber\\
&= -\,2\,\mu_{aA}\,e^{-i\,\pi\,\eta_{aA}^{bs}/2}\,\frac{1}{L_{l_{B}}^{C(+)}(i\,\kappa_{aA}^{B})}\,\int\limits_{0}^{R_{aA}}\,{\rm d}r_{aA}\,r_{aA}\,\varphi_{l_{B}}^{C}(i\,\kappa_{aA}^{B}\,r_{aA})\,\Big({\overleftarrow {\hat T}}_{r_{aA}} - {\overrightarrow {\hat T}}_{r_{aA}} \Big)\,I_{aA\,\,{l_B}\,{j_B}\,J_{B}}^B({r_{aA}}).
\label{ANCwronskian1}
\end{align}
Taking into account that 
\begin{align}
f(x)\,\Big(\frac{{\overleftarrow d}^{2}}{dx^{2}}    - \frac{{\overrightarrow d}^{2}}{dx^{2}} \Big)\,g(x\, 
= \frac{d}{dx}\,\Big(g(x)\,\frac{df(x)}{dx} - f(x)\,\frac{dg(x)}{dx} \Big)
\label{greentheorem1}
\end{align}
we arrive at the final expression for the ANC in terms of the Wronskian:
\begin{align}
C_{aA\,\,{l_B}\,{j_B}\,J_{B}}^{B}= \,e^{-i\,\pi\,\eta_{aA}^{bs}/2}\,\frac{1}{L_{l_{B}}^{C(+)}(i\,\kappa_{aA}^{B})}\,W[I_{aA\,\,{l_B}\,{j_B}\,J_{B}}^B({r_{aA}}),\,\varphi_{l_{B}}^{C}(i\,\kappa_{aA}^{B}\,r_{aA})]\Big|_{r_{aA}=R_{aA}},
\label{ANCwronskian2}
\end{align}
where the Wronskian 
\begin{align}
&W_{a}(i\,\kappa_{aA}^{B}\,r_{aA})\,=\,W[\,I_{aA\,\,{l_B}\,{j_B}\,J_{B}}^B({r_{aA}}),\,\varphi_{l_{B}}^{C}(i\,\kappa_{aA}^{B}\,r_{aA})]                                           \nonumber\\
&= I_{aA\,\,{l_B}\,{j_B}\,J_{B}}^B({r_{aA}})\,\frac{{\rm d}\varphi_{l_{B}}^{C}(i\,\kappa_{aA}^{B}\,r_{aA})\,}{{\rm d}r_{aA}}   - \varphi_{l_{B}}^{C}(i\,\kappa_{aA}^{B}\,r_{aA})\,\,\frac{{\rm d}I_{aA\,\,{l_B}\,{j_B}\,J_{B}}^B({r_{aA}})}{{\rm d}r_{aA}} .
\label{wronskian1}
\end{align} 
It is straightforward to see that determined here ANC is a real quantity because $\varphi_{l_{B}}^{C}(i\,\kappa_{aA}^{B}\,r_{aA})$ given by Eq. (\ref{varphiFl1}), $\,I_{aA\,\,{l_B}\,{j_B}\,J_{B}}^B({r_{aA}})$  and 
$\,e^{-i\,\pi\,\eta_{aA}^{bs}/2}\,\frac{1}{L_{l_{B}}^{C(+)}(i\,\kappa_{aA}^{B})}$ are real at pure imaginary momentum $k_{aA}= i\,\kappa_{aA}^{B}$.  
Correspondingly, 
the Coulomb renormalized ANC is given by
\begin{align}
{\tilde C}_{aA\,\,{l_B}\,{j_B}\,J_{B}}^{B}= (2\,\kappa_{aA}^{B})^{l_{B}}\,\frac{\Gamma(l_{B} + 1)}{\Gamma(2\,l_{B} + 2)}\,W[I_{aA\,\,{l_B}\,{j_B}\,J_{B}}^B({r_{aA}}),\,\varphi_{l_{B}}^{C}(i\,\kappa_{aA}^{B}\,r_{aA})]\Big|_{r_{aA}=R_{aA}}.
\label{CoulrenormANCwronskian1}
\end{align}

We know that the Wronskian calculated for two independent solutions of the Schr\"odinger equation is constant \cite{newton}. Because the radial overlap function $ I_{aA\,\,{l_B}\,{j_B}\,J_{B}}^B({r_{aA}})$ is not a solution of the Schr\"odinger equation in the nuclear interior, the Wronskian and, hence, the ANC determined by Eq. (\ref{ANCwronskian1}) depend on the channel radius $R_{aA}$, if it is not too large. However, if the adopted channel radius is large enough, we can replace the radial overlap function by its asymptotic term, see Eq. (\ref{asymptoverlap1}), proportional to the Whittaker function, which determines the radial shape of the asymptotic radial overlap function and is a singular solution of the radial 
Schr\"odinger equation. Thus, because $\varphi_{l_{B}}^{C}(i\,\kappa_{aA}^{B}\,r_{aA})$ is independent regular solution of the same equation, at large $R_{aA}$, where nuclear $a-A$ interaction can be neglected and asymptotic Eq. (\ref{asymptoverlap1}) can be applied, taking into account that $W[\,f_{l_{B}}^{C(+)}(i\,\kappa_{aA}^{B},r_{aA}),\,f_{l_{B}}^{C(-)}(i\,\kappa_{aA}^{B},r_{aA})]= 2\,\kappa_{aA}^{B}$ and Eq. (\ref{varphiFl1}) we get at large $R_{aA}$
\begin{align}
\,e^{-i\,\pi\,\eta_{aA}^{bs}/2}\,\frac{1}{L_{l_{B}}^{C(+)}(i\,\kappa_{aA}^{B})}\,W[I_{aA\,\,{l_B}\,{j_B}\,J_{B}}^B({r_{aA}}),\,\varphi_{l_{B}}^{C}(i\,\kappa_{aA}^{B}\,r_{aA})]\Big|_{r_{aA}=R_{aA}}
=C_{aA\,\,{l_B}\,{j_B}\,J_{B}}^{B}.
\label{wronskianANC1}
\end{align}  
Hence Eq. (\ref{ANCwronskian2}) at large $R_{aA}$, as expected, turns into identity and proof of it is an additional test that Eq. (\ref{ANCwronskian2}) is correct. However our idea is to use Eq. (\ref{ANCwronskian1}) at $R_{aA}$, which doesn't
exceed the radius of nucleus $B=(aA)$. In the nuclear interior the contemporary microscopic models provide quite accurate overlap functions.
The ANC calculated using Eq. (\ref{ANCwronskian1}) may depend on the adopted channel radius $R_{aA}$ but the ratio of the mirror ANCs, as it is shown below, practically is not sensitive to $R_{aA}$. It allows us to analyze the impact of the Coulomb effects on the ANC by separating different scales of these effects.

\section{Comparison of mirror proton and neutron ANCs and Coulomb effects}  

In this section we compare the mirror proton and neutron ANCs and analyze the Coulomb effects 
which are responsible for the difference between these ANCs. In what follows we simplify the notations for the proton and neutron ANCs omitting the quantum numbers and using just $C_{p}$ and $C_{n}$, correspondingly. From Eq. (\ref{ANCwronskian2}) 
we get the ratio of the squares of the proton and neutron ANCs for mirror nuclei:
\begin{align}
L_{pn}^{W}(R_{NA}) = \frac{C_{p}^{2}}{C_{n}^{2}}\, =\, \Big[\Big(\frac{\kappa_{pA}^{B}}{\kappa_{nA}^{A+1}}\Big)^{l_{B}}\,\frac{\Gamma(l_{B} + 1 + \eta_{pA}^{bs})}{\Gamma(l_{B} + 1)}\,\frac{W_{p}(i\,\kappa_{pA}^{B}\,R_{NA})}{W_{n}(i\,\kappa_{nA}^{A+1}\,R_{NA})}\Big]^{2},
\label{ratioANCpn1}
\end{align}
where $W_{p}(i\,\kappa_{pA}^{B}\,R_{NA})$ $\,\big(W_{n}(i\,\kappa_{nA}^{A+1}\,R_{NA}) \big)$  is the Wronskian calculated for the proton (neutron) at the channel radius $R_{NA}$. To get the proton (neutron) Wronskian we can just replace in Eq. (\ref{wronskian1}) $\,a \to p\,$ ($a \to n$).
For the neutron Wronskian $\varphi_{l_{B}}(i\,\kappa_{nA}^{A+1}\,r_{nA})= \,r_{nA}^{l_{B}+1}\,
e^{-\kappa_{nA}^{A+1}\,r_{nA}}\,{}_{1}F_{1}(l_{B}+1, 2\,l_{B} +2;\,2\,\kappa_{nA}^{A+1}\,r_{nA})$, where $\kappa_{nA}^{A+1}=\sqrt{2\,\mu_{nA}\,\varepsilon_{nA}^{A+1}}$ is the wave number of the bound state $A+1=(n\,A)$ of the isobaric analogue state of the bound state $B=(p\,A)$.

Calculation of the ratio of the mirror nucleon ANCs requires knowledge of the microscopic radial overlap functions. Meantime in \cite{timofeyuk2003} another expression for the mirror nucleon ANCs ratio was obtained which can be used when the overlap functions are not available. I will show here how simple this derivation when using Eq. (\ref{ratioANCpn1}).
First, as it was pointed out in \cite{timofeyuk2003}, in the nuclear interior Coulomb interaction varies very little in the vicinity of $R_{NA}$ and its effect leads only to shifting of the nucleon binding energy.  Hence, we assume that $\varphi_{l_{B}}^{C}(i\,\kappa_{aA}^{B}\,r_{aA})$ and $\varphi_{l_{B}}(i\,\kappa_{aA}^{B}\,r_{aA})$ behave similarly at $r_{NA} \approx R_{NA}$ except for the overall normalization, that is 
\begin{align}
\varphi_{l_{B}}^{C}(i\,\kappa_{pA}^{B}\,r_{pA})= \frac{\varphi_{l_{B}}^{C}(i\,\kappa_{pA}^{B}\,R_{NA})}{\varphi_{l_{B}}(i\,\kappa_{nA}^{A+1}\,R_{NA})}\,\varphi_{l_{B}}(i\,\kappa_{nA}^{A+1}\,r_{pA}).
\label{varphipn1}
\end{align}
Then Eq. (\ref{ratioANCpn1}) reduces to 
\begin{align}
L^{W\,'}_{pn}(R_{NA}) = \frac{C_{p}^{2}}{C_{n}^{2}}\, \approx\, \Bigg[\Big(\frac{\kappa_{pA}^{B}}{\kappa_{nA}^{A+1}}\Big)^{l_{B}}\,\frac{\Gamma(l_{B} + 1 + \eta_{pA}^{bs})}{\Gamma(l_{B} + 1)}\,\frac{\varphi_{l_{B}}^{C}(i\,\kappa_{pA}^{B}\,R_{NA})}{\varphi_{l_{B}}(i\,\kappa_{nA}^{A+1}\,R_{NA})}\,\frac{{W[I_{pA{\kern 1pt} {\kern 1pt} {l_B}{\kern 1pt} {j_B}{\kern 1pt} {J_B}}^B({r_{pA}}){\mkern 1mu},\,\varphi _{{l_B}}(i{\mkern 1mu} \kappa _{pA}^B{\mkern 1mu} {r_{pA}})]{\Big|_{{r_{pA}} = {R_{NA}}}}}}{{W[I_{nA{\kern 1pt} {\kern 1pt} {l_B}{\kern 1pt} {j_B}{\kern 1pt} {J_B}}^{A+1}({r_{nA}}){\mkern 1mu},\, \varphi _{{l_B}}(i{\mkern 1mu} \kappa _{nA}^{A + 1}{\mkern 1mu} {r_{nA}})]{\Big|_{{r_{nA}} = {R_{NA}}}}}}\Bigg]^{2}.
\label{ratioANCpnapp1}
\end{align}
Neglecting further the difference between the proton and neutron mirror overlap functions in the nuclear interior we obtain the approximate ratio of the squares of the mirror ANCs from \cite{timofeyuk2003} (in the notations of the current paper):
\begin{align}
L_{pn}(R_{NA}) = \frac{C_{p}^{2}}{C_{n}^{2}}\, \approx  \Big[\Big(\frac{\kappa_{pA}^{B}}{\kappa_{nA}^{A+1}}\Big)^{l_{B}}\,\frac{\Gamma(l_{B} + 1 + \eta_{pA}^{bs})}{\Gamma(l_{B} + 1)}\,\frac{\varphi_{l_{B}}^{C}(i\,\kappa_{pA}^{B}\,R_{NA})}{\varphi_{l_{B}}(i\,\kappa_{nA}^{A+1}\,R_{NA})}\,\Big]^{2}.
\label{ratioANCpnapp2}
\end{align}
In descending accuracy I can rank Eq. (\ref{ratioANCpn1}) as the most accurate, then Eq. (\ref{ratioANCpnapp1}) and  then  Eq. (\ref{ratioANCpnapp2}). Taking into account that the microscopic overlap functions (calculated in no-core-shell-model \cite{navratil2000,navratil2003,quaglioni}, variational Monte Carlo method \cite{pieper,nollett2007,nollett2011} or oscillator shell-model \cite{tim2011}) are more accurate in the nuclear interior, using Eq (\ref{ratioANCpn1}) one can determine the ratio of the mirror ANCs quite accurately.   
Then follows Eq. (\ref{ratioANCpnapp1}), in which only one approximation (\ref{varphipn1}) is used, and finally  Eq. (\ref{ratioANCpnapp2}), in which two approximations are used. Eq. (\ref{ratioANCpn1}) requires knowledge of two mirror overlap nucleon functions, Eq. (\ref{ratioANCpnapp1}) requires only one overlap functions while to calculateEq. (\ref{ratioANCpnapp2}) one doesn't need overlap functions and it is the simplest to estimate. As we will see below, while ratio (\ref{ratioANCpn1}) practically doesn't depend on the channel radius $R_{NA}$, ratios (\ref{ratioANCpnapp1}) and (\ref{ratioANCpnapp2}) show some evident dependence on $R_{NA}$. 
   
Now we explain physics causing the difference between the proton and neutron mirror ANCs. 
First of all the proton ANC is affected by the main CRF ${\Rcu}_{1}$, see Eq. (\ref{CRF1}). Eliminating this major factor from the proton ANC we obtain the ratio of the squares of the Coulomb renormalized proton and neutron ANCs:
\begin{align}
{\tilde L}_{pn}^{W}(R_{NA}) = \frac{{\tilde C}_{p}^{2}}{C_{n}^{2}}\, =\, \Big[\Big(\frac{\kappa_{pA}^{B}}{\kappa_{nA}^{A+1}}\Big)^{l_{B}}\,\frac{W_{p}(i\,\kappa_{pA}^{B}\,R_{NA})}{W_{n}(i\,\kappa_{nA}^{A+1}\,R_{NA})}\Big]^{2},
\label{tilderatioANCpn1}
\end{align}
\begin{align}
{\tilde L}_{pn}^{W\,'}(R_{NA}) = \frac{{\tilde C}_{p}^{2}}{C_{n}^{2}}\, \approx\, \Bigg[\Big(\frac{\kappa_{pA}^{B}}{\kappa_{nA}^{A+1}}\Big)^{l_{B}}\,\frac{\varphi_{l_{B}}^{C}(i\,\kappa_{pA}^{B}\,R_{NA})}{\varphi_{l_{B}}(i\,\kappa_{nA}^{A+1}\,R_{NA})}\,\frac{{W[I_{pA{\kern 1pt} {\kern 1pt} {l_B}{\kern 1pt} {j_B}{\kern 1pt} {J_B}}^B({r_{pA}}){\mkern 1mu},\,\varphi _{{l_B}}(i{\mkern 1mu} \kappa _{pA}^B{\mkern 1mu} {r_{pA}})]{\Big|_{{r_{pA}} = {R_{NA}}}}}}{{W[I_{nA{\kern 1pt} {\kern 1pt} {l_B}{\kern 1pt} {j_B}{\kern 1pt} {J_B}}^{A+1}({r_{nA}}){\mkern 1mu},\, \varphi _{{l_B}}(i{\mkern 1mu} \kappa _{nA}^{A + 1}{\mkern 1mu} {r_{nA}})]{\Big|_{{r_{nA}} = {R_{NA}}}}}}\Bigg]^{2}
\label{tilderatioANCpnapp1}
\end{align}
and
\begin{align}
{\tilde L}_{pn} = \frac{{\tilde C}_{p}^{2}}{C_{n}^{2}}\, \approx  \Big[\Big(\frac{\kappa_{pA}^{B}}{\kappa_{nA}^{A+1}}\Big)^{l_{B}}\,\frac{\varphi_{l_{B}}^{C}(i\,\kappa_{pA}^{B}\,R_{NA})}{\varphi_{l_{B}}(i\,\kappa_{nA}^{A+1}\,R_{NA})}\,\Big]^{2}.
\label{tildeANCratiotim1}
\end{align}

These ratios are still may be far from unity. After we removed the main CRF, there are two more remaining CRFs, which determine the difference between the proton and neutron mirror ANCs. The second factor appears because of the difference in the proton and neutron binding energies. The dependence of the ANC on the binding energy is exponential \cite{baz}.
Because the binding energy of the neutron analogue state is larger then the corresponding proton binding energy, the renormalized squared proton ANC ${\tilde C}_{p}^{2}$ is lower then $C_{n}^{2}$, and the second CRF decreases the proton ANC: the bigger the difference $\varepsilon_{nA}^{A+1} - \varepsilon_{pA}^{B}$ the stronger decrease of the proton ANC. The third CRF, which increases the proton ANC compared to the neutron one, is generated by the fine Coulomb effects ( the effects left after removal the main CRF and difference in the binding energies) and is minor compared to the first two CRFs. Now after this discussion we can rewrite Eq. (\ref{tilderatioANCpn1}) as
\begin{align}
{\tilde L}_{pn}^{W}(R_{NA}) = {\tilde {\Rcu}}_{2}^{W}\,{\tilde {\Rcu}}_{3}^{W},
\label{tilderatioANCpn3}
\end{align}
where
\begin{align}
{\tilde {\Rcu}}_{2}^{W}= \Big[\Big(\frac{\kappa_{pA}^{B}}{\kappa_{nA}^{A+1}}\Big)^{l_{B}}\,\frac{W_{n}(i\,\kappa_{pA}^{B}\,R_{NA})}{W_{n}(i\,\kappa_{nA}^{A+1}\,R_{NA})}\Big]^{2}
\label{bindenwr1}
\end{align}
and 
\begin{align}
{\tilde {\Rcu}}_{3}^{W}= \Big|\frac{W_{p}(i\,\kappa_{pA}^{B}\,R_{NA})}{W_{n}(i\,\kappa_{pA}^{B}\,R_{NA})} \Big|^{2}.
\label{fineCouleffwr1}
\end{align}
To calculate the binding energy effect ${\tilde {\Rcu}}_{2}^{W}$ on the ANCs ratio it is enough to replace in Eq. (\ref{tilderatioANCpn1}) the proton Wronskian $W_{p}(i\,\kappa_{pA}^{B}\,R_{NA})$ by the neutron one $W_{n}(i\,\kappa_{pA}^{B}\,R_{NA})$ but calculated at the proton binding energy. To calculate the impact of the fine Coulomb effects it is enough to consider the ratio of the squared
proton and neutron Wronskians both calculated at the proton binding energy. Similarly we can estimate these renormalization effects for ${\tilde L}^{W\,'}_{pn}(R_{NA})$. For the ratio of the ANCs  ${\tilde L}_{pn}$ considered in \cite{timofeyuk2003}  we get
\begin{align}
{\tilde L}_{pn} = {\tilde {\Rcu}}_{2}\,{\tilde {\Rcu}}_{3},
\label{ANCratiotim2}
\end{align}
where 
\begin{align}
{\tilde {\Rcu}}_{2}= \Big[\Big(\frac{\kappa_{pA}^{B}}{\kappa_{nA}^{A+1}}\Big)^{l_{B}}\,\frac{\varphi_{l_{B}}(i\,\kappa_{pA}^{B}\,R_{NA})}{\varphi_{l_{B}}(i\,\kappa_{nA}^{A+1}\,R_{NA})}\,\Big]^{2}
\label{bindenrgeff1}
\end{align}
is the CRF determining the effect of the binding energy, while
\begin{align}
{\tilde {\Rcu}}_{3}= \Big[\frac{\varphi_{l_{B}}^{C}(i\,\kappa_{pA}^{B}\,R_{NA})}{\varphi_{l_{B}}(i\,\kappa_{pA}^{B}\,R_{NA})}\,\Big]^{2}
\label{fineCouleff1}
\end{align}
is the CRF determining the fine Coulomb effects.

Thus we can express the squared proton ANC in terms of the mirror squared neutron one as 
\begin{align}
C_{p}^{2}= {\Rcu}_{1}\,{\tilde {\Rcu}}_{2}^{W}\,{\tilde {\Rcu}}_{3}^{W}\,
C_{n}^{2}
\label{pANCnANCrelatwr1}
\end{align}
or
\begin{align}
C_{p}^{2}= {\Rcu}_{1}\,{\tilde {\Rcu}}_{2}\,{\tilde {\Rcu}}_{3}\,
C_{n}^{2},
\label{pANCnANCrelat1}
\end{align}
depending on whether we use the Wronskian approach or more simple renormalization from \cite{timofeyuk2003}. Here we took into account that the quantum numbers of the proton and neutron analogue states are the same. 

Because I don't have all the required overlap functions, which are necessary to calculate the CRFs, only the renormalization factors ${\tilde {\Rcu}}_{2}$ and ${\tilde {\Rcu}}_{3}$ are estimated in Sect. \ref{calculations1}. 

\section{Calculations}
\label{calculations1}

In this section we consider some cases applying both Wronskian formalism and Eqs. (\ref{ratioANCpnapp1}) and (\ref{ratioANCpnapp2}).  

\subsubsection{Comparison of ANCs for ${}^{41}{\rm Sc}(1f_{7/2};\,0.0\, {\rm MeV}) \to {}^{40}{\rm Ca}(0.0\, {\rm MeV}) + p$ and ${}^{41}{\rm Ca}(1f_{7/2};\, 0.0\, {\rm MeV}) \to {}^{40}{\rm Ca}(0.0\, {\rm MeV}) + n$.}
I start from analysis of the mirror proton and neutron ANCs for ${}^{41}{\rm Sc}(1f_{7/2};\,0.0\,{\rm MeV}) \to {}^{40}{\rm Ca}(0.0\,{\rm MeV}) + p$ and ${}^{41}{\rm Ca}(1f_{7/2};\, 0.0\, {\rm MeV}) \to {}^{40}{\rm Ca}(0.0\, {\rm MeV}) + n$, correspondingly.  Both ANCs correspond to isobaric analogue states in the mirror nuclei and we can apply the formalism discussed in the previous section. The overlap functions are taken from \cite{tim2011}.

\begin{figure}
\epsfig{file=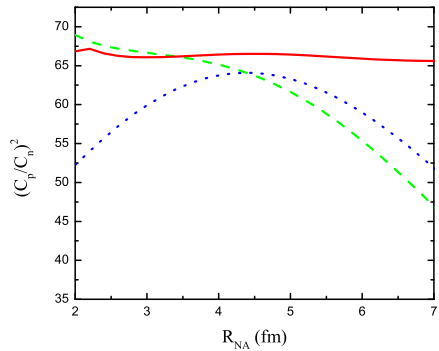,width=7cm}
\caption{(Color online)
Ratio of the square of the proton ANC for the virtual decay ${}^{41}{\rm Sc}(1f_{7/2};\,0.0\, {\rm MeV}) \to {}^{40}{\rm Ca}(0.0\,{\rm MeV}) + p$ to the square of the neutron ANC for the mirror nucleus virtual decay ${}^{41}{\rm Ca}(1f_{7/2};\,0.0\, {\rm MeV}) \to {}^{40}{\rm Ca}(0.0\,{\rm MeV}) + n$;
solid red line - Wronskian method, Eq. (\ref{ratioANCpn1}), green dashed line is obtained using Eq. (\ref{ratioANCpnapp1}) and dotted blue line is obtained using Eq. (\ref{ratioANCpnapp2}) as in \cite{timofeyuk2003}.} 
\label{fig_41Ca41ScANCratio}
\end{figure}

In Fig. \ref{fig_41Ca41ScANCratio} is shown the ratio of the ANCs calculated using the Wronskian formalism and approximations (\ref{ratioANCpnapp1}) and (\ref{ratioANCpnapp2}). While the ratio obtained in the Wronskian formalism remains practically constant, both approximated ratios depend on the channel radius $R_{NA}$. Note that the results obtained using approximations (\ref{ratioANCpnapp1}) and (\ref{ratioANCpnapp2}) are valid only at $R_{NA} \leq  R_{A}$, where $R_{A} \approx 4.5 $ fm is the radius of $A= {}^{40}{\rm Ca}$. Eq. (\ref{ratioANCpnapp1}) better agrees with the exact Wronskian expression than Eq. (\ref{ratioANCpnapp2}). The latter has the smallest deviation from the exact result at $R_{NA} \approx 4.4$.    

\begin{figure}
\epsfig{file= 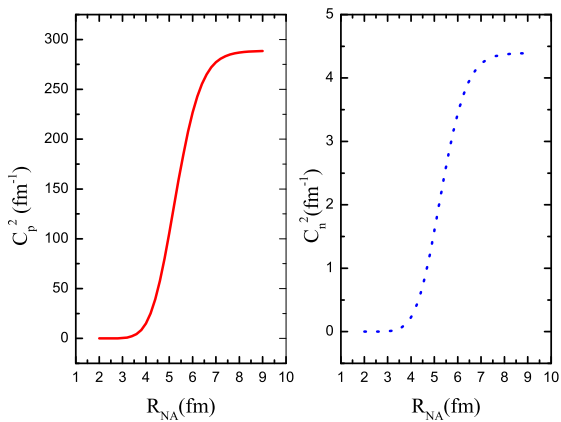,width=8cm}
\caption{(Color online) Panel (a): solid red line- dependence on $R_{NA}$ of the square of the proton ANC for the virtual decay ${}^{41}{\rm Sc}(1f_{7/2};\,0.0\,{\rm MeV}) \to {}^{40}{\rm Ca}(0.0\, {\rm MeV}) + p$ 
calculated using the Wronskian expression given by Eq. (\ref{ANCwronskian2});
panel (b): blue dotted line- dependence on $R_{NA}$ of the square of the neutron ANC for the virtual decay ${}^{41}{\rm Ca}(1f_{7/2};\,0.0\, {\rm MeV}) \to {}^{40}{\rm Ca}(0.0\, {\rm MeV}) + n$ 
calculated using the Wronskian expression given by Eq. (\ref{ANCwronskian2}).} 
\label{fig_CpCn(41Sc41Ca)(gr)}
\end{figure}

In Fig. \ref{fig_CpCn(41Sc41Ca)(gr)} the dependence on $R_{NA}$ of the square of the proton and neutron ANCs for the virtual decays ${}^{41}{\rm Sc}(1f_{7/2};\,0.0\,{\rm MeV}) \to {}^{40}{\rm Ca}(0.0\,{\rm MeV}) + p$ and ${}^{41}{\rm Ca}(1f_{7/2};\,0.0\,{\rm MeV}) \to {}^{40}{\rm Ca}(0.0\,{\rm MeV}) + n$ 
calculated using the Wronskian expression given by Eq. (\ref{ANCwronskian1}). 
Comparison of Figs \ref{fig_41Ca41ScANCratio} and \ref{fig_CpCn(41Sc41Ca)(gr)} is very instructive. 
As we can conclude from the latter the ANCs calculated using the Wronskian method reach their correct values at $R_{NA} > 7$ fm. In other words, only at $R_{NA} > 7$ fm the proton and neutron overlap functions can be replaced by their asymptotic terms. However, the ratio of the ANCs calculated using the Wronskian formalism remains constant practically at all $R_{nA} >2$, allowing one to calculate this ratio even in the nuclear interior where the overlap functions cannot be replaced by their asymptotic terms.
The calculated using the Wronskian method the proton and neutron ANCs and theirs ratio are in excellent agreement with the calculations in \cite{tim2011}. 
 
Now we consider how the Coulomb renormalization affects the proton ANC $C_{p}^{2}= 286.9$ fm${}^{-1}$.
There are three CRFs. Dividing the the proton squared ANC by each of these factors we can eliminate step by step all three Coulomb effects eventually arriving at the neutron squared ANC. To estimate the different CRFs I use Eqs (\ref{bindenrgeff1}) and (\ref{fineCouleff1}). The Wronskian formalism aslo could be used but not all the needed overlap functions are available. The main CRF in the case under consideration ${\Rcu}_{1}= 14311.9$. Hence the Coulomb renormalized square of the proton ANC is ${\tilde C}_{p}^{2}= C_{p}^{2}/{\Rcu}_{1}= 0.02$ fm${}^{-1}$. 
Now we take into account the remaining two CRFs coming from difference in the proton and neutron binding energies and the residual Coulomb effects. Because at $R_{NA}=4.4$ fm the ratio of the proton and neutron ANCs obtained using Eq. (\ref{ratioANCpnapp2}) is the closest to the one obtained by the Wronskian method, the remaining two Coulomb renormalizations are calculated at $R_{NA}=4.45$ fm. 
The binding energy of the proton in $\,{}^{41}{\rm Sc}(0.0\,{\rm MeV})\,$ is  $\,\varepsilon_{p\,{}^{40}{\rm Ca}}^{{}^{41}{\rm Sc}}= 1.085$ MeV, while the neutron binding energy in $\,{}^{41}{\rm Ca}(0.0\,{\rm MeV})$ is $\,\varepsilon_{n\,{}^{40}{\rm Ca}}^{{}^{41}{\rm Ca}}= 8.362$ MeV. This large difference in the neutron and proton binding energies leads to significant renormalization of the proton ANC compared to the neutron one. Dividing ${\tilde C}_{p}^{2}$ by the CRF ${\Rcu}_{2}= 0.0011$ we get the renormalized proton ANC at the neutron binding energy: ${\tilde C}_{p}^{'\,2}= 18.18$ fm${}^{-1}$. Finally, dividing it by the CRF ${\Rcu}_{3}= 4.16$, reflecting the residual Coulomb effects, we get the square of the neutron ANC $C_{n}^{2}= 4.37$ fm${}^{-1}$.  These calculations clearly demonstrate the scale of the different CRFs leading to the difference between the mirror proton and neutron ANCs. 

\begin{figure}
\epsfig{file= 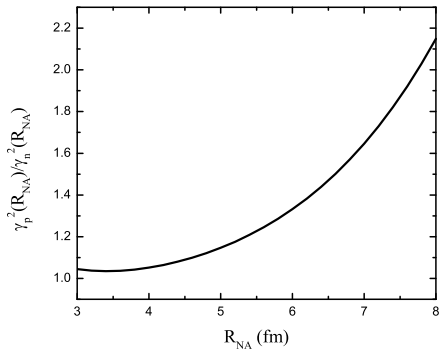,width=7cm}
\caption{(The dependence on the channel radius of the ratio of the proton and neutron reduced widths for ${}^{41}{\rm Sc}(1f_{7/2};\,0.0\, {\rm MeV}) \to {}^{40}{\rm Ca}(0.0\, {\rm MeV}) + p$ and ${}^{41}{\rm Ca}(1f_{7/2};\,0.0\, {\rm MeV}) \to {}^{40}{\rm Ca}(0.0\, {\rm MeV}) + n$, correspondingly.  The nucleon ANCs  calculated using the Wronskian expression given by Eq. (\ref{ANCwronskian2}).} 
\label{fig_redwidthratio41Sc41Ca}
\end{figure}

Let us consider now the ratio of the proton and neutron reduced widths for the mirror nuclei:
\begin{align}
\frac{\gamma _{p}^2}{\gamma_{n}^{2}} = \frac{{W_{ - {\eta _{p}^{bs}},\,{l_B}\, + \,1/2}^2(2{\kappa _{p}\,{R_{NA}})\,}C_{p}^2}}{{W_{ 0,\,{l_B}\, + \,1/2}^2(2{\kappa _{n}\,{R_{NA}})\,}\,C_{n}^2}}.
\label{redwidthratio1}
\end{align}
For the case under consideration this ratio is $\gamma _{p}^2/\gamma _{n}^2= 1.1$ at $R_{NA}=4.4 $ fm
and gradually increasing with $R_{NA}$ increase, see Fig \ref{fig_redwidthratio41Sc41Ca}, demonstrating model dependence of the reduced widths. Because the ratio of the nucleon ANCs remains constant, the channel radius dependence comes entirely from the channel radius dependence of the Whittaker functions.  The higher binding energy of the neutron generates stronger channel radius dependence of
the neutron Whittaker function, which is actually proportional to the spherical Hankel function. 

\subsubsection{Comparison of ANCs for ${}^{17}{\rm F}(1d_{5/2}) \to {}^{16}{\rm O}(0.0\,{\rm MeV}) + p$ and ${}^{17}{\rm O}(1d_{5/2};\, 0.0\, {\rm MeV}) \to {}^{16}{\rm O}(0.0\,{\rm MeV}) + n$.}
\label{17F17Od521}

As in the previous case, the overlap functions are nodeless at $r_{NA} >0$ and I will demonstrate how the Wronskian method works compared to approximated equations. I start from the analysis of the proton and neutron ANCs for ${}^{17}{\rm F}(1d_{5/2};\,0.0\, {\rm MeV}) \to {}^{16}{\rm O}(0.0\,{\rm MeV}) + p$ and ${}^{17}{\rm O}(1d_{5/2};\,0.0 {\rm MeV}) \to {}^{16}{\rm O}(0.0\, {\rm MeV}) + n$, correspondingly. Both ANCs correspond to isobaric analogue states in mirror nuclei and we can apply the formalism discussed in the previous section. The overlap functions are taken from \cite{tim2011}.

\begin{figure}
\epsfig{file=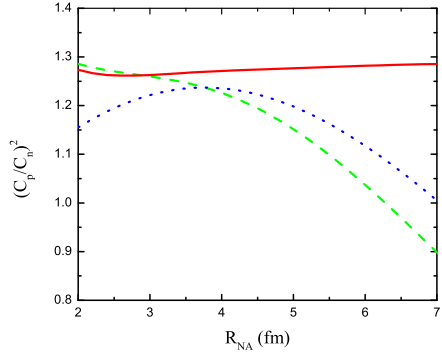,width=7cm}
\caption{(Color online) Ratio of the square of the proton ANC for the virtual decay ${}^{17}{\rm F}(1d_{5/2};\,0.0\,{\rm MeV}) \to {}^{16}{\rm O}(0.0\, {\rm MeV}) + p$ to the square of the neutron ANC for the mirror nucleus virtual decay ${}^{17}{\rm O}(1d_{5/2};\,0.0\,{\rm MeV}) \to {}^{16}{\rm O}(0.0\, {\rm MeV}) + n$;
solid red line - the Wronskian method, Eq. (\ref{ratioANCpn1}), green dashed line is obtained using Eq. (\ref{ratioANCpnapp1}) and dotted blue line is obtained using Eq. (\ref{ratioANCpnapp2}) as in \cite{timofeyuk2003}.} 
\label{fig_17F17OANCratiod52}
\end{figure}

\begin{figure}
\epsfig{file=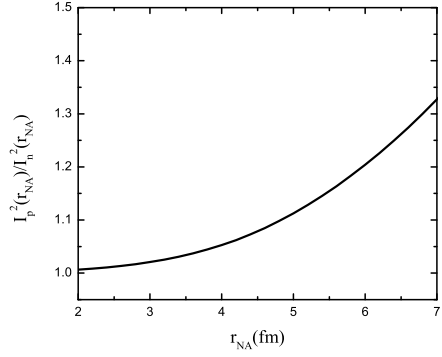,width=7cm}
\caption{
Ratio of the square of the proton and neutron radial overlap functions for ${}^{17}{\rm F}(1d_{5/2};\,0.0\,{\rm MeV}) \to {}^{16}{\rm O}(0.0\, {\rm MeV}) + p$ and ${}^{17}{\rm O}(1d_{5/2};\,0.0\,{\rm MeV}) \to {}^{16}{\rm O}(0.0\, {\rm MeV}) + n$. The overlap functions are from \cite{tim2011}.} 
\label{fig_overlapratio17F17Od52}
\end{figure}

In Fig. \ref{fig_17F17OANCratiod52} the dependence on $R_{NA}$ of the square of the ratio of the mirror proton and neutron ANCs for the virtual decays ${}^{17}{\rm F}(1d_{5/2}) \to {}^{16}{\rm O}(0.0\,{\rm MeV}) + p$ and ${}^{17}{\rm O}(1d_{5/2}) \to {}^{16}{\rm O}(0.0 {\rm MeV}) + n$ calculated using the Wronskian formalism and Eqs. (\ref{ratioANCpnapp1}) and (\ref{ratioANCpnapp2}) is shown. While the ratio obtained in the Wronskian formalism remains practically constant, Eqs. (\ref{ratioANCpnapp1}) and (\ref{ratioANCpnapp2}) provide the ratios which depend on the channel radius $R_{NA}$. Eq. (\ref{ratioANCpnapp1}) agrees very well with the exact Wronskian method at $R_{NA} \leq 3$ fm. Assumption that the mirror proton and neutron overlap functions are close to each other in the nuclear interior is the basis for both approximate equations for the ANCs ratio. It is clear from Fig. \ref{fig_overlapratio17F17Od52} that with the $R_{NA}$ increase the square of the ratio of the mirror overlap functions deviates from unity.   

\begin{figure}
\epsfig{file= 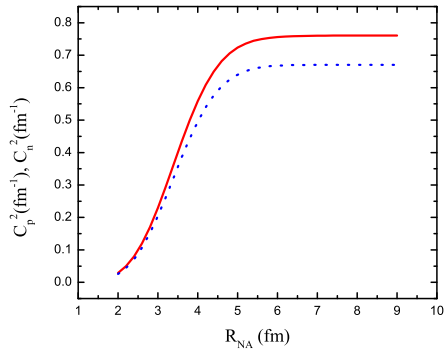,width=7cm}
\caption{ (Color online) Dependence on $R_{NA}$ of the square of the proton (solid red line) and neutron (blue dotted line) ANCs for the virtual decays ${}^{17}{\rm F}(1d_{5/2}) \to {}^{16}{\rm O}(0.0\,{\rm MeV}) + p$ and ${}^{17}{\rm O}(1d_{5/2}) \to {}^{16}{\rm O}(0.0\,{\rm MeV}) + n$, correspondingly, calculated using the Wronskian expression given by Eq. (\ref{ANCwronskian2}).} 
\label{fig_CpCn(17F17O)(gr)}
\end{figure}
Comparison of Figs \ref{fig_17F17OANCratiod52} and \ref{fig_CpCn(17F17O)(gr)} is very instructive and shows the power of the Wronskian method. 
As it is clear from Fig. \ref{fig_CpCn(17F17O)(gr)}, the proton and neutron ANCs calculated using the Wronskian equation reach their plateau only at $R_{NA} > 6$ fm. However, the ratio of the ANCs calculated using the Wronskian formalism,  Eq. (\ref{ratioANCpn1}), remains constant practically at all $R_{nA} >2$ fm, allowing one to calculate this ratio even in the region where the overlap functions cannot be replaced by their asymptotic terms. The ratio of the proton and neutron ANCs calculated
using approximated  Eq. (\ref{ratioANCpnapp1}) agrees very well with the Wronskian method at $R_{nA} \leq 3$ fm, that is well in the nuclear interior, while Eq. (\ref{ratioANCpnapp2}) gives the ratio below the Wronskian method.  

Now we consider how the Coulomb renormalization affects the proton ANC $C_{p}^{2}= 0.58$ fm${}^{-1}$. To estimate the different CRFs I use, as in the previous case, Eqs (\ref{bindenrgeff1}) and (\ref{fineCouleff1}).  The main CRF in the case under consideration ${\Rcu}_{1}= 42.01$. Hence the Coulomb renormalized square of the proton ANC is ${\tilde C}_{p}^{2}= C_{p}^{2}/{\Rcu}_{1}= 0.0138$ fm${}^{-1}$. 
Now we take into account the remaining two CRFs coming from difference in the proton and neutron binding energies and the residual Coulomb effects. Because at $R_{NA}=4.0$ fm the ratio of the proton and neutron ANCs obtained using Eq. (\ref{ratioANCpnapp2}) is the closest to the one obtained by the Wronskian method, the remaining two CRFs are calculated at $R_{NA}=4.0$ fm. 
The binding energy of the proton in $\,{}^{17}{\rm F}(0.0\,{\rm MeV})\,$ is $\,\varepsilon_{p\,{}^{16}{\rm O}}^{{}^{17}{\rm F}}= 0.605$ MeV, while the neutron binding energy in $\,{}^{17}{\rm O}(0.0\,{\rm MeV})$ is $\,\varepsilon_{n\,{}^{16}{\rm O}}^{{}^{17}{\rm O}}= 4.14$ MeV. This large difference in the neutron and proton binding energies leads to significant renormalization of the proton ANC compared to the neutron one. Dividing ${\tilde C}_{p}^{2}$ by the CRF ${\Rcu}_{2}= 0.015$ we get the renormalized proton ANC at the neutron binding energy: ${\tilde C}_{p}^{'\,2}= 0.92$ fm${}^{-1}$. Finally, dividing it by the CRF ${\Rcu}_{3}= 1.97$, reflecting the residual Coulomb effects, we get the square of the neutron ANC $C_{n}^{2}= 0.467$ fm${}^{-1}$.  These calculations once again demonstrate the scale of the different CRFs leading to the difference between the mirror proton and neutron ANCs.

\begin{figure}
\epsfig{file= 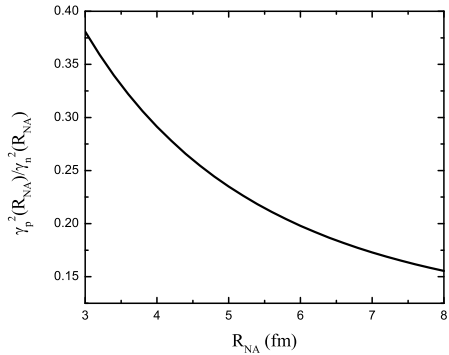,width=7cm}
\caption{Dependence on the channel radius of the ratio of the proton and neutron reduced widths for ${}^{17}{\rm F}(1d_{5/2};\,0.0\, {\rm MeV}) \to {}^{16}{\rm O}(0.0\, {\rm MeV}) + p$ and ${}^{17}{\rm O}(1d_{5/2};\,0.0\, {\rm MeV}) \to {}^{16}{\rm O}(0.0\,{\rm MeV}) + n$, correspondingly.  The nucleon ANCs calculated using the Wronskian expression given by Eq. (\ref{ANCwronskian2}).} 
\label{fig_redwidthratio17F17Od52}
\end{figure}

The ratio of the proton and neutron reduced widths $\gamma _{p}^2/\gamma _{n}^2= 0.29$ at $R_{NA}=4.0 $ fm
and gradually decreasing with $R_{NA}$ increase, see Fig \ref{fig_redwidthratio17F17Od52}, demonstrating model dependence of the reduced widths. In this case the proton Whittaker function drops faster than the neutron one.  

\subsubsection{Comparison of ANCs for ${}^{17}{\rm F}(2s_{1/2}) \to {}^{16}{\rm O}(0.0\,{\rm MeV}) + p$ and ${}^{17}{\rm O}(2s_{1/2}) \to {}^{16}{\rm O}(0.0\,{\rm MeV}) + n$.}

\begin{figure}
\epsfig{file=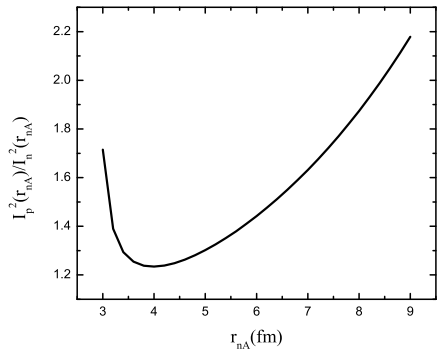,width=7cm}
\caption{
Ratio of the square of the proton and neutron radial overlap functions for ${}^{17}{\rm F}(2s_{1/2}) \to {}^{16}{\rm O}(0.0 {\rm MeV}) + p$ and ${}^{17}{\rm O}(2s_{1/2}) \to {}^{16}{\rm O}(0.0 {\rm MeV}) + n$.} 
\label{fig_overlapratio17F17Os0.5}
\end{figure}

In this section I analyze the ratio of the proton and neutron ANCs for ${}^{17}{\rm F}(2s_{1/2}) \to {}^{16}{\rm O}(0.0 {\rm MeV}) + p$ and ${}^{17}{\rm O}(2s_{1/2}) \to {}^{16}{\rm O}(0.0 {\rm MeV}) + n$, correspondingly.  Both ANCs correspond to isobaric analogue states in the mirror nuclei and we can apply the formalism discussed in the previous section. The overlap functions are taken from \cite{tim2011}. It is a special case because of the very low proton binding energy, $\varepsilon_{p\,{}^{16}{\rm O}}^{{}^{17}{\rm F}}= 0.105$ MeV. Besides, the overlap functions have one node at $r_{NA} >0$ and it will be interesting to see how the Wronskian method works in this case. The ratio of the square of the proton and neutron overlap functions  $I_{p}^{2}(r_{NA})/I_{n}^{2}(r_{NA})>1$ at all $r_{NA}$, see Fig \ref{fig_overlapratio17F17Os0.5}. Hence both approximations (\ref{ratioANCpnapp1}) and (\ref{ratioANCpnapp2}) 
fail.

\begin{figure}
\epsfig{file=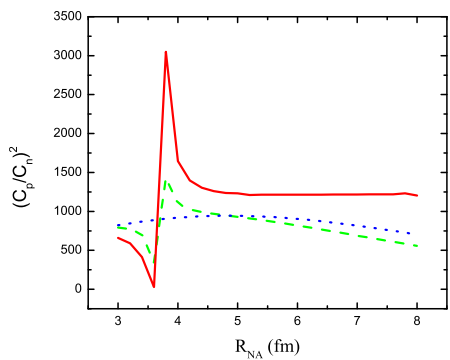,width=7cm}
\caption{(Color online) Ratio of the square of the proton ANC for the virtual decay ${}^{17}{\rm F}(2s_{1/2}) \to {}^{16}{\rm O}(0.0 {\rm MeV}) + p$ to the square of the neutron ANC for the mirror nucleus virtual decay ${}^{17}{\rm O}(2s_{1/2}) \to {}^{16}{\rm O}(0.0 {\rm MeV}) + n$;
solid red line - the Wronskian method, Eq. (\ref{ratioANCpn1}), green dashed line is obtained using Eq. (\ref{ratioANCpnapp1}) and dotted blue line is obtained using Eq. (\ref{ratioANCpnapp2}) as in \cite{timofeyuk2003}.} 
\label{fig_17F17OANCratio}
\end{figure}

It is seen from Fig. \ref{fig_17F17OANCratio} where the ratios of the ANCs calculated using the Wronskian formalism and Eqs. (\ref{ratioANCpnapp1}) and (\ref{ratioANCpnapp2}) are shown. While the ratio obtained in the Wronskian formalism remains practically constant at $R_{NA} > 4$ fm, the ratios obtained using Eqs. (\ref{ratioANCpnapp1}) and (\ref{ratioANCpnapp2}) depend on the channel radius $R_{NA}$ and at the best radius $R_{NA}= 5$ fm underestimate the exact ratio (Wronskian equation) by 30\%. 
The anomalous behavior of the Wronskian expression at $R_{NA} < 4$ fm is the result of the nodes of the overlap functions. 

\begin{figure}
\epsfig{file= 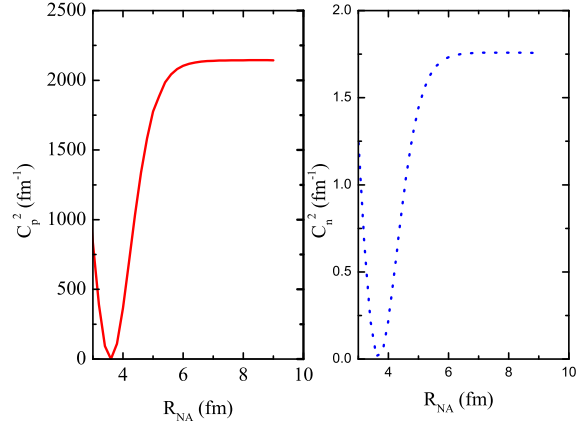,width=8cm}
\caption{ (Color online) Panel (a): red solid line- dependence on $R_{NA}$ of the square of the proton ANC for the virtual decay ${}^{17}{\rm F}(2s_{1/2}) \to {}^{16}{\rm O}(0.0 {\rm MeV}) + p$
calculated using the Wronskian expression given by Eq. (\ref{ANCwronskian2});
panel (b): blue dotted line -  dependence on $R_{NA}$ of the square of the neutron ANC for the virtual decay ${}^{17}{\rm O}(2s_{1/2}) \to {}^{16}{\rm O}(0.0 {\rm MeV}) + n$ 
calculated using the Wronskian expression given by Eq. (\ref{ANCwronskian2}).} 
\label{fig_CpCn(17F17O)(s1/2)}
\end{figure}

In Fig. \ref{fig_CpCn(17F17O)(s1/2)} the dependence on $R_{NA}$ of the square of the mirror proton and neutron ANCs for the virtual decays ${}^{17}{\rm F}(2s_{1/2}) \to {}^{16}{\rm O}(0.0\,{\rm MeV}) + p$ and ${}^{17}{\rm O}(2s_{1/2}) \to {}^{16}{\rm O}(0.0\,{\rm MeV}) + n$ calculated using the Wronskian expression given by Eq. (\ref{ANCwronskian2}). 
Owing to the low nucleon binding energies the overlap functions the nucleon overlap functions reach their asymptotic tail (the ANCs reach their plateau) only at $R_{NA} > 7$ fm, see Fig. \ref{fig_CpCn(17F17O)(s1/2)}. However, the ratio of the ANCs calculated using the Wronskian formalism, see Fig. \ref{fig_17F17OANCratio},   remains practically constant at all $R_{nA} >5$ fm, allowing one to calculate this ratio even in the region where the overlap functions cannot be replaced by their asymptotic terms.  Note that the calculated using the Wronskian method the proton and neutron ANCs and theirs ratio are in excellent agreement with the calculations in \cite{tim2011}. 
 
\begin{figure}
\epsfig{file= 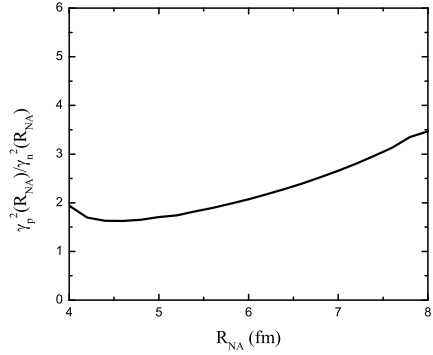,width=7cm}
\caption{The dependence on the channel radius of the ratio of the proton and neutron reduced widths for ${}^{17}{\rm F}(2s_{1/2};\,0.0\, {\rm MeV}) \to {}^{16}{\rm O}(0.0\,{\rm MeV}) + p$ and ${}^{17}{\rm O}(2s_{1/2};\,0.0\, {\rm MeV}) \to {}^{16}{\rm O}(0.0\,{\rm MeV}) + n$, correspondingly.  The nucleon ANCs calculated using the Wronskian expression given by Eq. (\ref{ANCwronskian2}).} 
\label{fig_redwidthratio17F17Os0.5}
\end{figure}

The ratio of the proton and neutron reduced widths $\gamma _{p}^2/\gamma _{n}^2= 1.71$ at $R_{NA}=5.0 $ fm
and gradually increasing with $R_{NA}$ increase, see Fig \ref{fig_redwidthratio17F17Os0.5}, demonstrating model dependence of the reduced widths. In this case the neutron Whittaker function drops slightly faster than the proton one.  

\subsubsection{Comparison of ANCs for ${}^{8}{\rm B}(1p_{3/2};\,0.0\, {\rm MeV}) \to {}^{7}{\rm Be}(0.0\, {\rm MeV}) + p$ and ${}^{8}{\rm Li}(1p_{3/2};\,0.0\, {\rm MeV}) \to {}^{7}{\rm Li}(0.0\, {\rm MeV}) + n$.}

To analyze the mirror proton and neutron ANCs for ${}^{8}{\rm B}(1p_{3/2};\,0.0\, {\rm MeV}) \to {}^{7}{\rm Be}(0.0 {\rm MeV}) + p$ and ${}^{8}{\rm Li}(1p_{3/2};\,0.0\, {\rm MeV}) \to {}^{7}{\rm Li}(0.0\, {\rm MeV}) + n$  I use the overlap functions obtained from the variational Monte Carlo wave functions using Green's function method \cite{nollett2012}.  

\begin{figure}
\epsfig{file=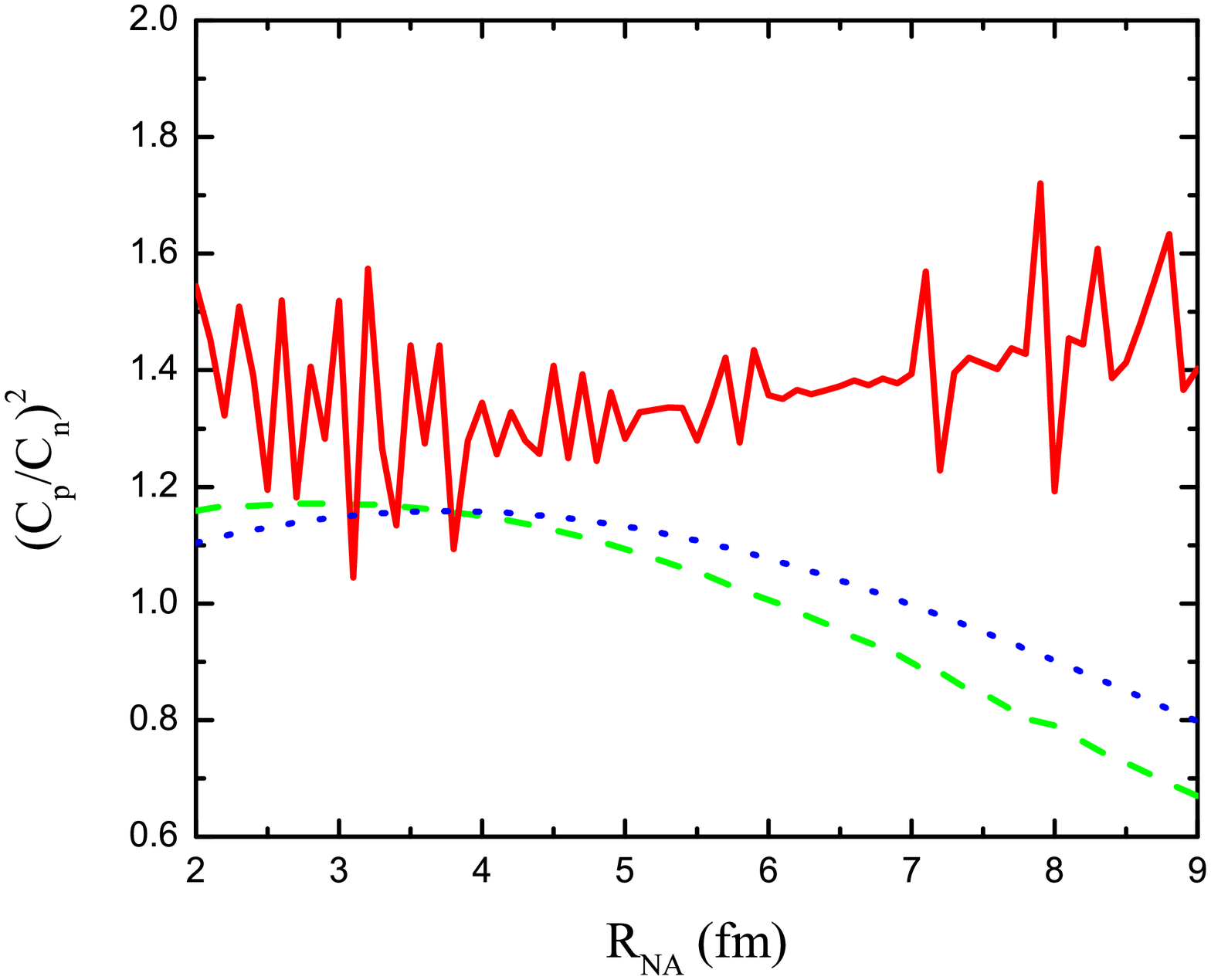,width=7cm}
\caption{(Color online) Ratio of the square of the proton ANC for the virtual decay ${}^{8}{\rm B}(1p_{32/2};\,0.0\,{\rm MeV}) \to {}^{7}{\rm Be}(0.0\,{\rm MeV}) + p$ to the square of the neutron ANC for the mirror nucleus virtual decay ${}^{8}{\rm Li}(1p_{3/2};\,0.0\,{\rm MeV}) \to {}^{7}{\rm Li}(0.0\,{\rm MeV}) + n$;
solid red line - the Wronskian method, Eq. (\ref{ratioANCpn1}), green dashed line is obtained using Eq. (\ref{ratioANCpnapp1}) and dotted blue line is obtained using Eq. (\ref{ratioANCpnapp2}).} 
\label{fig_8B8LiANCratio}
\end{figure}

In Fig. \ref{fig_8B8LiANCratio} the ratios of the square of the proton and neutron ANCs calculated using the Wronskian formalism and Eqs. (\ref{ratioANCpnapp1}) and (\ref{ratioANCpnapp2}) are shown. The ratio obtained in the Wronskian formalism remains almost constant ( the observed oscillations are related with accuracy of the Monte Carlo method).  The ratios obtained using Eqs. (\ref{ratioANCpnapp1}) and (\ref{ratioANCpnapp2}) in the internal region at $R_{NA} < 3.5$ fm  practically constant and lower than the exact ratio. The mean value of the calculated ratio of the square of the proton and neutron ANCs using the Wronskian method in the interval $3.4-4.6$ fm is $1.24$, what is slightly higher then the experimental ratio $1.06 \pm 0.11$ \cite{trache2003} and the ratio $1.15$ obtained using Eq. (\ref{ratioANCpnapp2}). 

\begin{figure}
\epsfig{file=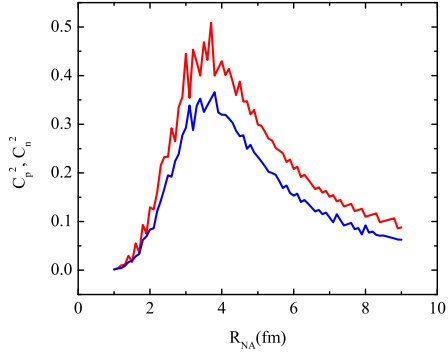,width=7cm}
\caption{(Color online) Red (blue) solid line: dependence on $R_{NA}$ of the square of the proton (neutron) ANC for the virtual decay ${}^{8}{\rm B}(1p_{32/2};\,0.0\,{\rm MeV}) \to {}^{7}{\rm Be}(0.0\,{\rm MeV}) + p\,$ \Big(${}^{8}{\rm Li}(1p_{3/2};\,0.0\,{\rm MeV}) \to {}^{7}{\rm Li}(0.0\,{\rm MeV}) + n$ \Big)
calculated using the Wronskian expression given by Eq. (\ref{ANCwronskian2}).} 
\label{fig_CpCn7Bep7Lin(p32)}
\end{figure}

In Fig. \ref{fig_CpCn7Bep7Lin(p32)} is shown the dependence on $R_{NA}$ of the square of the proton and neutron ANCs for the virtual decays ${}^{8}{\rm B}(1p_{32/2};\,0.0\,{\rm MeV}) \to {}^{7}{\rm Be}(0.0\,{\rm MeV}) + p$ and ${}^{8}{\rm Li}(1p_{3/2}; 0.0\,{\rm MeV}) \to {}^{7}{\rm Li}(0.0\,{\rm MeV}) + n$, correspondingly,
calculated using the Wronskian expression given by Eq. (\ref{ANCwronskian2}). 
As we can see, in contrast to the previous cases, the calculated using the Wronskian approach ANCs never reach plateau, what reflects the fact that the Monte Carlo overlap functions don't have correct asymptotic radial shape.  Neverhteless, the ratio of the mirror proton and neutron ANCs, as it is shown in Fig. \ref{fig_8B8LiANCratio}, remains practically stable confirming once more that this ratio can be calculated with the overlap functions, which are correct only in the nuclear interior. 

Once again we can estimate all the Coulomb renormalization effects on the proton ANC. 
We adopt the square of the ANC $C_{p}^{2}= 0.43$ fm${}^{-1}$ obtained from the Wronskian expression at $R_{NA}=4$ fm. The main CRF in the case under consideration ${\Rcu}_{1}= 13.66$. The Coulomb renormalized square of the proton ANC is ${\tilde C}_{p}^{2}= 0.0315$ fm${}^{-1}$. Now we take into account the remaining two Coulomb renormalizations coming from difference in the proton and neutron binding energies and the residual Coulomb effects.  
The binding energy of the proton in ${}^{8}{\rm B}(0.0\,{\rm MeV})$ is $\varepsilon_{p\,{}^{7}{\rm Be}}^{{}^{8}{\rm B}}= 0.1375$ MeV, while the neutron binding energy in ${}^{8}{\rm Li}(0.0\,{\rm MeV})$ is $\varepsilon_{n\,{}^{7}{\rm Li}}^{{}^{8}{\rm Li}}= 2.03$ MeV. The difference between the proton and neutron binding energies leads to renormalization of the proton ANC compared to the neutron one. Dividing ${\tilde C}_{p}^{2}$ by the CRF at $R_{NA}=4$ fm $\,{\Rcu}_{2}= 0.053$ we get the renormalized proton ANC at the neutron binding energy: $\,{{\tilde C}_{p}}^{'\,2}= 0.59$ fm${}^{-1}$. Finally, dividing it by the CRF at $R_{NA}=4\,$ fm $\,{\Rcu}_{3}= 1.61\,$ reflecting the residual Coulomb effects we get the square of the neutron ANC $\,C_{n}^{2}= 0.37$ fm${}^{-1}$.  

\begin{figure}
\epsfig{file= 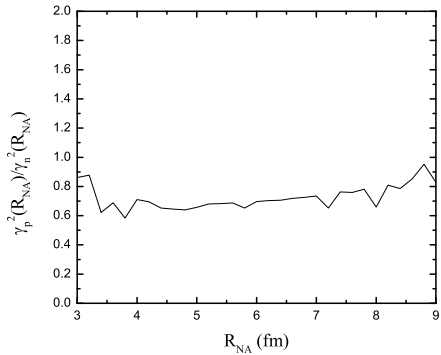,width=7cm}
\caption{Dependence on the channel radius of the ratio of the proton and neutron reduced widths for ${}^{8}{\rm B}(1p_{3/2};\,0.0\, {\rm MeV}) \to {}^{7}{\rm Be}(0.0\, {\rm MeV}) + p$ and ${}^{8}{\rm Li}(1p_{3/2};\,0.0\, {\rm MeV}) \to {}^{7}{\rm Li}(0.0\, {\rm MeV}) + n$, correspondingly.  The nucleon ANCs  calculated using the Wronskian expression given by Eq. (\ref{ANCwronskian2}).} 
\label{fig_redwidthratio8B8Lip32}
\end{figure}

The ratio of the proton and neutron reduced widths for ${}^{8}{\rm B}(1p_{3/2};\,0.0\, {\rm MeV}) \to {}^{7}{\rm Be}(0.0\, {\rm MeV}) + p$ and ${}^{8}{\rm Li}(1p_{3/2};\,0.0\, {\rm MeV}) \to {}^{7}{\rm Li}(0.0\, {\rm MeV}) + n$, correspondingly, see Fig \ref{fig_redwidthratio8B8Lip32}, gradually increases with $R_{NA}$ increase. 

\section{Summary}

The first goal of this paper is to analyze the Coulomb renormalization of the ANC. It is shown that
the Coulomb renormalization of the proton ANC compared to the neutron one of the mirror nucleus consists of three factors: the main CRF given by Eq. (\ref{CRF1}). The second CRF is the result of the decrease of the proton binding energy compared to the neutron one and the thrid 
CRF is the result of the fine Coulomb effects (residual Coulomb effects after removing the main CRF).
The scale of each CRF is determined for different cases. I also draw attention on the possibility of using the 
renormalized ANC when the standard one becomes too large.

The second important result of this paper is derivation of new expression for the ratio of the proton and neutron ANCs for isobaric analogue states of the mirror nuclei. This equation is obtained from the Pinkston-Satchler equation and given by the ratio of the Wronskians taken from the radial overlap functions and regular solutions
of the two-body $N + A$ radial Schr\"odinger equation with the short-range interaction excluded. It is shown that using microscopic overlap functions, which are more accurate in the nuclear interior, one can determine the ratio of the mirror nucleon ANCs which are the amplitudes of the tails of the corresponding overlap functions. It allows us to obtain one of the nucleon ANCs if other is known. 
The results of this paper can be extended for the resonance states (when the overlap functions for resonance states will be available) and for the arbitrary mirror ANCs for the systems $a+ A$, for example for $a=alpha$.  

\section{acknowledgments}  
The work was supported by the US Department of Energy under Grants No. DE-FG02-93ER40773, No. DE-FG52-
09NA29467, No. DE-SC0004958 (topical collaboration TORUS) and NSF under Grant No. PHY-0852653. The author expresses his thanks to N. K. Timofeyuk, K. M. Nollett and R. B. Wiringa for presenting the overlap functions.


\begin{thebibliography}{99}
\bibitem{blokh77} L. D. Blokhintsev, I. Borbely, and E. I. Dolinskii, Fiz. Elem. Chastits t. Yadra {\bf 8}, 1189 (1977) [Sov. J. Part. Nucl. {\bf 8}, 485 (1977)].
\bibitem{blokhintsev84} L. D. Blokhintsev,  A. M. Mukhamedzhanov, A. N. Safronov, Fiz. Elem. Chastits t. Yadra {\bf 15}, 1296 (1984) [Sov. J. Part. Nucl. {\bf 15}, 580 (1984)].
\bibitem{muktr99}  A. M. Mukhamedzhanov and R. E. Tribble, Phys. Rev. {\bf C 59}, 3418 (1999).
\bibitem{muk77} E. I. Dolinsky and A. M. Mukhamedzhanov,  Izv. N. SSSR, Ser. Fiz. {\bf 41}, 2055 (1977) [Bull. cad. Sci. USSR, Phys. Ser. {\bf 41}, 55 (1977)].
\bibitem{muktim90} A. M. Mukhamedzhanov and N. K. Timofeyuk, Pis'ma Eksp. Teor. Fiz. {\bf 51}, 247 (1990) [JETP Lett. {\bf 51}, 282 (1990)].
\bibitem{muk90} A. M. Mukhamedzhanov and N. K. Timofeyuk, Sov. J. Nucl. Phys. 51, 431 (1990) [Yad. Fiz. 51, 679 (1990)]. 
\bibitem{muk2001}   A. M. Mukhamedzhanov, C. A. Gagliardi, and R. E. Tribble, Phys. Rev. {\bf C 63}, 024612 (2001) 
\bibitem{muk2011} A. M. Mukhamedzhanov, L. D. Blokhintsev, and B.  F. Irgaziev, Phys. Rev. {\bf C 83}, 055805 (2011). 
\bibitem{tang} Xiaodong Tang, A. Azhari, C. A. Gagliardi, A. M. Mukhamedzhanov, F. Pirlepesov, L. Trache, R. E. Tribble, V. Burjan, V. Kroha, and F. Carstoiu, Phys. Rev. {\bf C 67}, 015804 (2003). 
\bibitem{kroha2011} A. M. Mukhamedzhanov, M. La Cognata and V. Kroha,  Phys. Rev. {\bf C 83}, 044604 (2011).
\bibitem{kramers} H. A. Kramers, Hand und Jahrbuch der Chemischer  Physik {\bf 1}, 312 (1938).
\bibitem{perelomov}  M. Perelomov, V.S. Popov, M. V. Terent'ev,   ZhETF {\bf 51}, 309 (1966).  
\bibitem{heisenberg} W. Heisenberg, Zs f. Naturforsch. {\bf 1}, 608 (1946). 
\bibitem{meller} C. M\"oller, Dan. Vid Selsk. Mat. Fys. Medd. {\bf 22}, N.19 (1946). 
\bibitem{hu} N. Hu, Phys. Rev. {\bf 74}, 131 (1948). 
\bibitem{zeldovich} Ya. B. Zel'dovich, ZhetF  {\bf 51}, 1492 (1965).
\bibitem{hu94} H. M. Xu, C. A. Gagliardi, R. E. Tribble, A. M. Mukhamedzhanov and N. K. Timofeyuk, Phys. Rev. Lett. {\bf 73}, 2027 (1994).
\bibitem{muk95} A. M. Mukhamedzhanov, R. P. Schmitt, R. E. Tribble, and A. Sattarov, Phys. Rev. {\bf C 52}, 3483 (1995).
\bibitem{muk2003} A. M. Mukhamedzhanov {\it et al.}, Phys. Rev. {\bf C 67}, 065804  (2003).
\bibitem{muk2006} A. M. Mukhamedzhanov {\it et al.}, Phys. Rev. {\bf C 73}, 035806 (2006).
\bibitem{banu} A. Banu {\it et. al}, Phys. Rev. {\bf C 79}, 025805 (2009).
\bibitem{abdullah} T. Al-Abdullah {\it et. al}, Phys. Rev.  {\bf C 81}, 035802 (2010).
\bibitem{muk08} A. M. Mukhamedzhanov {\it et al.}, Phys. Rev. {\bf C 78}, 015804 (2008).
\bibitem{adelberger} E. G. Adelberger {\it et al.}, Rev. Mod. Phys. {\bf 83}, 195 (2011).
\bibitem{muk2010} A. M. Mukhamedzhanov and A. S. Kadyrov, Phys. Rev. {\bf C 82}, 051601(R) (2010). 
\bibitem{mukstripping2011} A. M. Mukhamedzhanov, Phys. Rev. {\bf C 84}, 044616 (2011).
\bibitem{timofeyuk2003} N. K. Timofeyuk, R. C. Johnson, and A. M. Mukhamedzhanov, Phys. Rev. Lett. {\bf 91}, 232501  (2003).
\bibitem{tim2005} N. K. Timofeyuk and P. Descouvemont, Phys. Rev. {\bf C 72}, 064324 (2005).
\bibitem{tim2006} N. K. Timofeyuk, D. Baye, P. Descouvemont, R. Kamouni, and I. J. Thompson,
Phys. Rev. Lett. {\bf 96}, 162501 (2006). 
\bibitem{nazarewicz} J. Okolowicz, N. Michel, W. Nazarewicz, and M. Ploszajczak, Phys. Rev {\bf C 85}, 064320 (2012).
\bibitem{titus} L. J. Titus, P. Capel, and F. M. Nunes, Phys. Rev. {\bf C 84}, 035805 (2011). 
\bibitem{johnson2006} E. D. Johnson {\it et al.}, Phys. Rev. Lett. {\bf 97}, 192701 (2006). 
\bibitem{pinkston} W. T. Pinkston and G. R. Satchler, Nucl. Phys. {\bf 72}, 641 (1965) 
\bibitem{philpott} R. J. Philpott, W. T. Pinkston, and G. R. Satchler, Nucl. Phys. {\bf A119}, 241 (1968).
\bibitem{wong} D. Y. Wong and H. P. Noyes, Phys. Rev. {\bf 126}, 1866 (1962).
\bibitem{mentkovsky} Yu. L. Mentkovsky, Nucl. Phys. {\bf 65}, 673 (1965). 
\bibitem{hamilton} J. Hamilton {\it et al.}, Nucl. Phys. {\bf 60}, 443 (1973).
\bibitem{baz} A. I. Baz', Ya. B. Zel'dovich, and A. M. Perelomov, Scattering, reactions and decay in nonrelativistic quantum mechanics, 2-nd edition, Nauka, Moscow (1971) [English translation of 1-st edition, Jerusalem (1969)]
\bibitem{blokhintsev83} L. D. Blokhintsev and A. N. Safronov, Izv. cad. Nauk SSSR {\bf 47}, 2168 (1983)
\bibitem{tim2011} N. K. Timofeyuk, Phys. Rev. {\bf C 84}, 054313 (2011)  
\bibitem{tim1998} N.K. Timofeyuk, Nucl. Phys. {\bf A632}, 19 (1998)
\bibitem{newton} R. G. Newton, Scattering Theory of Waves and Particles, 2nd ed., Springer-Verlag, Heidelberg, 1982.
\bibitem{navratil2000} P. Navratil, J. P. Vary and B. R. Barrett, Phys. Rev. {\bf C 62}, 054311 (2000)
\bibitem{navratil2003} P. Navratil, W. E. Ormand, Phys. Rev. {\bf C 68}, 034305 (2003)
\bibitem{quaglioni} S. Quaglioni, P. Navratil, Phys. Rev. Lett. {\bf 101}, 092501 (2008)  
\bibitem{pieper} C. S. Pieper, R. B. Wiringa,  Ann. Rev.  Nucl. Part. Sci. {\bf 51}, 53 (2001)
\bibitem{nollett2007} K. M. Nollett, S. C. Pieper, R. B. Wiringa, J. Carlson, G. M. Hale,  Phys. Rev. Lett.   {\bf 99}, 022502 (2007)
\bibitem{nollett2011} Kenneth M. Nollett and R. B. Wiringa, Phys. Rev {\bf C 83}, 041001 (2011).
\bibitem{jensen} O. Jensen, G. Hagen, M. Hjorth-Jensen, B. A. Brown, A. Gade, Phys. Rev. Lett. 107, 032501 (2011).
\bibitem{kadyrov2009} A. S. Kadyrov, I. Bray, A. M. Mukhamedzhanov and A. T. Stelbovics, Ann. Phys.  {\bf 324}, 1516 (2009)
\bibitem{strippingresonance2011}  A. M. Mukhamedzhanov, Phys. Rev. {\bf C 84}, 044616 (2011).
\bibitem{mukkadyrov}  A. M. Mukhamedzhanov and A. S. Kadyrov,  Phys. Rev. {\bf C 82}, 051601(R) (2010)
\bibitem{nollett2012} K. M. Nollett, arXiv:1206.0046 (2012).
\bibitem{trache2003} L. Trache, A. Azhari, F. Carstoiu, H. L. Clark, C. A. Gagliardi, Y.-W. Lui, A. M. Mukhamedzhanov, X. Tang, N. Timofeyuk, and R. E. Tribble, Phys. Rev. C 67, 062801(R) (2003)
\end{thebibliography}
\end{document}